\documentclass[preprint,12pt]{elsarticle}

\usepackage{amssymb}

\usepackage{hyperref}

\journal{SoSym Journal}
\usepackage[inline]{enumitem}
\usepackage{url}
\let\oldmarginpar\marginpar
\renewcommand{\marginpar}[1]{\leavevmode\oldmarginpar{\textit{{#1}}}}
\usepackage[parfill]{parskip}
\usepackage{multirow}
\usepackage{pdflscape}
\usepackage{xltabular}
\usepackage[table]{xcolor}
\usepackage{tabularx}
\usepackage{diagbox}
\usepackage{pgfplots}
\pgfplotsset{compat=1.16}
\usepackage{booktabs} 
\usepackage{caption} 
\makeatletter
\newcommand\footnoteref[1]{\protected@xdef\@thefnmark{\ref{#1}}\@footnotemark}
\makeatother

\usepackage{ifthen}
\usepackage{amssymb}
\usepackage{pifont}
\usepackage[normalem]{ulem} 
\definecolor{myGreen}{HTML}{009900}
\newcolumntype{L}{>{\raggedright\arraybackslash}m{8cm}}

\newboolean{showcomments}
\setboolean{showcomments}{false} 
\ifthenelse{\boolean{showcomments}}
  {
        \newcommand{\nb}[2]{
            \fcolorbox{gray}{yellow}{\bfseries\sffamily\scriptsize#1}
            {\sf\scriptsize$\blacktriangleright$\textit{{\sffamily\scriptsize#2}}$\blacktriangleleft$}
        }
        
        \newcommand{\nbc}[2]{
            \fcolorbox{gray}{orange}{\bfseries\sffamily\scriptsize#1}
            {\sf\scriptsize$\blacktriangleright$\textit{{\sffamily\scriptsize#2}}$\blacktriangleleft$}
        }        
		\newcommand{\nbb}[2]{
		\fcolorbox{black}{yellow}{\bfseries\scriptsize#1}
		{$\blacktriangleright$\textcolor{blue}{\textit{#2}}$\blacktriangleleft$}
		} 
		
        \newcommand{\nba}[2]{
            \fcolorbox{gray}{green}{\bfseries\sffamily\scriptsize#1}
            {\sf\small$\blacktriangleright$\textit{#2}$\blacktriangleleft$}
        }

        \newcommand{\nbz}[2]{
		\fcolorbox{white}{red}{\textcolor{white}{\bfseries\tiny#1}}
		{\textcolor{red}{\textit{#2}}}
	}

		\newcommand{\remarks}[1]{\color{red}[#1]\color{black}}

		\newcommand{\ins}[1]{\textcolor{blue}{\uline{#1}}} 
		\newcommand{\del}[1]{\textcolor{red}{\sout{#1}}} 
		\newcommand{\chg}[2]{\textcolor{red}{\sout{#1}}{\raw}\textcolor{blue}{\uline{#2}}} 
		\newcommand{\ugh}[1]{\textcolor{red}{\uwave{#1}}} 
            \newcommand\todo[1]{\nbz{!}{#1}}
            \newcommand\todoall[1]{\nbz{ToDo ALL}{#1}}

		\newcommand{\removed}[1]{\cbstart\removedfragile{#1}\cbend{}}
		\newcommand{\removedfragile}[1]{{\color{red}{\sout{#1}}}{}}

		\newcommand\simon[1]{\nba{Simon}{#1}}
            \newcommand\karo[1]{\nb{Karo}{#1}}
            \newcommand\stefanie[1]{\nb{Stefanie}{#1}}
            \newcommand\luca[1]{\nb{Luca}{#1}}
            \newcommand\tech[1]{}
            \newcommand\hiddencite[1]{}
            \newcommand\abbas[1]{\nbc{Abbas}{#1}}

  }
  {
		\newcommand{\nbb}[2]{}
		\newcommand{\remarks}[1]{}

		\newcommand{\ugh}[1]{#1} 
		\newcommand{\ins}[1]{#1} 
		\newcommand{\del}[1]{} 
		\newcommand{\chg}[2]{#2} 
		
		\newcommand{\removed}[1]{} 
        \newcommand{\removedfragile}[1]{}

        \newcommand\hiddencite[1]{}

	\newcommand\simon[1]{}
        \newcommand\karo[1]{}
        \newcommand\stefanie[1]{}
        \newcommand\luca[1]{}
        \newcommand\tech[1]{}        
        \newcommand\abbas[1]{}
        \newcommand\todo[1]{}
        \newcommand\todoall[1]{}
  }

\newcommand{\ext}[1]{}

\newcommand{\total}{18}

\begin{document}

\begin{frontmatter}




\title{Bridging MDE and AI: A Systematic Review of Domain-Specific Languages and Model-Driven Practices in AI Software Systems Engineering\tnoteref{t1}}

\tnotetext[t1]{This project has been partially supported and funded by the AIDOaRt project, an ECSEL Joint Undertaking (JU) under grant agreement No. 101007350 and the Austrian Research Promotion Agency (FFG) via the Austrian Competence Center for Digital Production (CDP) under the contract number 881843.}


\author[inst1,inst2]{Simon Rädler}
\author[inst3]{Luca Berardinelli}
\author[inst4]{Karolin Winter}
\author[inst3]{Abbas Rahimi}
\author[inst1]{Stefanie Rinderle-Ma}

\affiliation[inst1]{organization={Department of Computer Science, TUM School of Computation, Information and Technology, Technical University of Munich},
            addressline={Boltzmannstr. 3}, 
            city={Garching},
            postcode={85748}, 
            country={Germany}}

\affiliation[inst2]{organization={Business Informatics Group, Technical University of Vienna},
            addressline={Favoritenstraße 9-11/194-3}, 
            city={Vienna},
            postcode={1040}, 
            country={Austria}}

\affiliation[inst3]{organization={Department of Business Informatics–Software Engineering, Johannes Kepler University},
            addressline={Altenberger Straße 69}, 
            city={Linz},
            postcode={4040}, 
            country={Austria}}

\affiliation[inst4]{organization={Department of Industrial Engineering and Innovation Sciences, Eindhoven University of Technology},
            addressline={Groene Loper 3}, 
            city={Eindhoven},
            postcode={5612 AE}, 
            country={The Netherlands}}

\begin{abstract}
\textbf{Background:}
\luca{no acronyms in the abstract?}
Technical systems are becoming increasingly complex due to the increasing number of components, functions, and involvement of different disciplines.
In this regard, Model-Driven Engineering techniques and practices tame complexity during the development process by using models as primary artifacts. \ins{Modeling can be carried out through Domain-Specific Languages whose implementation is supported by model-driven techniques.}
Today, the amount of data generated during product development is rapidly growing, leading to an increased need to leverage Artificial Intelligence algorithms. However, using these algorithms in practice can be difficult and time-consuming. \chg{Therefore, utilizing MDE techniques and tools for formulating AI algorithms or parts of them can reduce these complexities and be advantageous.}{Therefore, leveraging domain-specific languages and model-driven techniques for formulating AI algorithms or parts of them can reduce these complexities and be advantageous.}

\textbf{Objective:}
This study aims to investigate the existing \chg{body of knowledge in the field of MDE in support of AI (MDE4AI)}{model-driven approaches relying on domain-specific languages in support of the engineering of AI software systems} to sharpen future research further and define the current state of the art.

\luca{check the numbers}
\textbf{Method:}
We conducted a Systemic Literature Review (SLR), collecting papers from five major databases resulting in 1335 candidate studies, eventually retaining 18 primary studies.
Each primary study will be evaluated and discussed with respect to the adoption of (1) MDE principles and practices and (2) the phases of AI development support aligned with the stages of the CRISP-DM methodology.

\textbf{Results:}
The study's findings show that language workbenches are of paramount importance in dealing with all aspects of modeling language development (metamodel, concrete syntax, and model transformation) and are leveraged to define domain-specific languages (DSL) explicitly addressing AI concerns.
The most prominent AI-related concerns are training and modeling of the AI algorithm, while minor emphasis is given to the time-consuming preparation of the data sets.
Early project phases that support interdisciplinary communication of requirements, such as the CRISP-DM \textit{Business Understanding} phase, are rarely reflected.

\textbf{Conclusion:}
The study found that the use of MDE for AI is still in its early stages, and there is no single tool or method that is widely used. Additionally, current approaches tend to focus on specific stages of development rather than providing support for the entire development process. As a result, the study suggests several research directions to further improve the use of MDE for AI and to guide future research in this area.

\end{abstract}




\end{frontmatter}


\section{Introduction}\label{sec:introduction}



\todo{Remove Arbiter \cite{zucker_arbiter_2020}}

\todo{Check what to do with \cite{delineGlindaSupportingData2021}}

Engineering systems are becoming more complex due to the increasing number of components, functions, and the involvement of several disciplines~\cite{beihoff_world_2014}.
To address this complexity, the integration of Model Driven Engineering (MDE) is promising~\cite{henderson_value_2021,huldt_state--practice_2019,madni_economic_2019}.

MDE aims to support the engineering of systems by providing and maintaining information for the development using models rather than documents~\cite{henderson_value_2021}.
Models are enriched with information, shared among stakeholders, and manipulated by Computer-Aided Software Engineering (CASE) tools, aiming at the highest possible degree of automation, e.g., via model transformations.
Consequently, MDE techniques allow informed decisions by producing and consuming models as machine-processable artifacts.

Although the models are enriched with information, MDE techniques lack the means to gather sufficient knowledge from more extensive data, e.g., Big Data or data collected from complex (software) systems.
In this respect, means of artificial intelligence (AI) and its sub-disciplines, namely machine learning (ML) and deep learning (DL), are beneficial to exploit the information hidden in data~\cite{provost_data_2013}.
Integrating data-driven methods to support engineering tasks has recently been defined as data-driven engineering~\cite{trauer_data-driven_2020}. It has proven to be beneficial in several engineering areas such as manufacturing~\cite{dogan_machine_2021}, aerospace industry~\cite{brunton_data-driven_2021} or other industrial applications~\cite{bertolini_machine_2021,fahle_systematic_2020,sarker_machine_2021}.

AI integration is mainly case-specific, and thus there are several methods supporting the implementation of AI in literature~\cite{azevedo_kdd_2008}, e.g., Cross Industry Standard Process for Data Mining (CRISP-DM)~\cite{wirth_crisp-dm_2000}, which support the development of AI tools by providing support for the specific development steps typically applied to AI projects.
The incorporation of methods such as CRISP-DM and MDE capabilities is rarely considered in the literature, even though the need for experts to implement data-driven solutions is increasing due to the requirement to integrate AI into existing methods~\cite{davey_systems_2022} and the implementation effort is decreasing by applying MDE principles and practices to the development of AI capabilities.

In recent times, a series of workshops has been initiated, focusing on the intersection of Model-Driven Engineering (MDE) techniques and Artificial Intelligence (AI)~\cite{burgueno_guest_2022,burgueno_mde_2021,burgueno_mde_2021-1}. Specifically, these workshops explore the use of MDE practices to define AI methods (referred to as MDE4AI) and to provide AI support for MDE (AI4MDE). The goal is to enhance software development by automating engineering activities through techniques like code generation. The current state of practice and state-of-the-art MDE approaches that facilitate the implementation of AI capabilities remain insufficiently explored. 


In this context, the overall Research Goal (RG) of our systematic study can be defined as presented in Table~\ref{tab:goal}. The definition of the RG aligns with the Goal-Question-Metric perspective, as proposed by Basili et al.~\cite{basili_goal_1994}.


\begin{table*}[htb]
\centering
\caption{The overall research goal}
\resizebox{15cm}{!}{%
    \begin{tabular}{p{2cm}|p{13cm}}
    \textit{Purpose} & Collection and comparison of studies on \\ 
    \textit{Issue} & model-driven approaches leveraging suitably designed domain-\newline specific languages that explicitly address the engineering of \\ 
    \textit{Object} & software systems leveraging AI/ML \\
    \textit{Viewpoint} & from the point of view of researchers. \\
    \end{tabular}
}
\label{tab:goal}
\end{table*}

We define the following overarching Research Question (RQ) based on this research goal:

\begin{enumerate}[align=left,label=\textbf{Main RQ}]
    \item What is the current state of the art for model-driven engineering \ugh{with extensions to formalize}\luca{?} artificial intelligence methods and applications?
\end{enumerate}


To address the main research question, various refined and more fine-grained RQs are introduced in Section~\ref{sec:method}.

\luca{The following paragraph have been rewritten}
\todo{Check if  "ii) the support to data mining steps." is the intended purpose of the CRISP DM part of this work}
\ins{For this reason, we conducted a systematic literature review according to the guidelines set out by~\cite{kitchenham_systematic_2013} to address the identified research questions~\cite{petersen_guidelines_2015} and spotlight model-driven approaches that leverage suitably designed DSLs to automate the engineering of systems with AI capabilities, specifically emphasizing MDE4AI, and ii) the support to data mining steps.}
\ugh{Accordingly, AI techniques for enhancing MDE approaches fall outside the scope of our work.}\luca{redundant? should we remove it?}


\luca{Do we use the term quantitative assessment criteria later in the paper?}
The contributions of this SLR comprise:
\begin{itemize}
    \item Collection and analysis of state-of-the-art model-driven approaches \ins{ adopting specialized DSLs for AI applications in (software) system engineering}.
    \item \del{Introduction of }Quantitative assessment criteria for model-driven approaches for AI.
    \item Quantitative assessment of existing approaches with derived research opportunities.
\end{itemize}

The remainder of the paper is organized as follows. Section~\ref{sec:background} provides background on terms such as MDE and AI.
Section~\ref{sec:method} introduces the research methodology, i.e., the paper search and selection process.
Section~\ref{sec:results} presents the approaches aligned with the data extraction strategy of the SLR protocol in Section~\ref{sec:method}.
Section~\ref{sec:discussion} answers the research questions, discusses the key findings, and depicts implications and future research.  Section~\ref{sec:threatstovalidity} assesses the quality and limitations of the current SLR using threats to validity analysis. Finally, 
Section~\ref{sec:conclusion} summarizes the paper.
\section{Background}\label{sec:background}
This section presents relevant background on MDE and AI.
The focus is on fundamentals that support understanding of the SLR results and is not intended to reflect the current state of the art in each research area. 

\subsection{Model-Driven Engineering}\label{sec:background_MDE}
The core of MDE includes the pillar concepts of \textit{model}, \textit{metamodel}, and \textit{model transformation}~\citep{brambilla_model-driven_2017,RODRIGUESDASILVA2015139}. 


\textit{Models} are machine-readable artifacts representing particular concerns of a system under study, such as design-time information like software architecture or hardware platform, or operational information like monitored data. 
\textit{Metamodels} define the modeling concepts and their relationships, providing an intentional description of all possible models that must conform to the associated metamodel. From a language engineering perspective, a metamodel represents the \textit{abstract syntax} of a modeling language. \textit{Metamodels} define modeling languages conceptually and are independent of any concrete representation. The \textit{concrete syntax} of a language assigns graphical or textual elements to metamodel elements that can be understood by users and edited through model editors~\cite{brambilla_model-driven_2017,RODRIGUESDASILVA2015139}.
As models in MDE are considered machine-readable artifacts, so-called \textit{model transformations} apply to modifying existing or generating new engineering artifacts. 
These artifacts are then used for particular purposes.\luca{added distinction between dev and runtime models} \ins{During the development process, models support} realizing the steps of the envisioned engineering process toward the (partial or full) generation of the software system.
\ins{At runtime, models provide intelligent support during execution. According to~\cite{bencomo2019}}, \textit{"a runtime model is defined as a casually connected self-representation of the associated system that emphasizes the structure, behaviour, and goal of the system and which can be manipulated at runtime for specific purposes"}.

Depending on the specific engineering concerns (software, hardware, or system as a whole) as well as the role played by model artifacts due to degrees of automation of model management activities (e.g., model-based being to refer to a lighter version of model-driven), different modeling acronyms are typically used (e.g., MBE, MBSE, MDE, MDSE. MDD, MDA)\footnote{See \url{https://modeling-languages.com/clarifying-concepts-mbe-vs-mde-vs-mdd-vs-mda/} for a discussion.}.
The various modeling acronyms show that the MDE community is widespread, and the same applies to the goals and applications of MDE approaches.

\ins{Recently, so-called low-code and no-code development platforms gained the attention of researchers and the market~\cite{diruscio2022}. LCDPs leverage MDE principles by utilizing automation, analysis, and abstraction through modeling and metamodeling~\cite{sahay2020}. One of the goals of these platforms is to offer degrees of automation in the generation of software applications while partially or completely hiding code to their users, typically referred to as citizen developers, i.e., users with limited programming experience in the software development process.}

\subsection{Artificial Intelligence}\label{sec:background_AI}
\textit{Artificial intelligence} is a flourishing science with numerous practical applications, ranging from image/voice recognition to recommendation systems and self-driving cars. The primary goal of AI is to tackle problems that are tough for humans but relatively simple for computers~\cite{GoodBengCour16}. In a non-scientific context, the different terms around AI are quite fuzzy, and depending on the application area, different terms like machine learning (ML), deep learning, data science, data mining, and so on are used exchangeable~\cite{emmert-streib_defining_2019}. In science, some of the terms are today used interchangeable, while in computer science, the terms are precisely defined. 
For example, \textit{data science} is defined as the umbrella term referring to the broad field of extracting information and knowledge by analyzing data to derive patterns, trends, etc., and report them as human-understandable insights~\cite{sanchez-pinto_big_2018} that are beneficial for various areas~\cite{shinde_review_2018}, such as manufacturing~\cite{dogan_machine_2021}.
Another sample is Data mining~\citep{provost_data_2013}, which is the extraction of knowledge from data sets or systems and processes.

Although each term refers to a specific subcategory of data science, the implementation phases are essentially similar. 
Therefore, in the following, all synonyms and sub-terms will be related to the term AI for better understanding.
Consequently, methodologies have been developed to structure and support the implementation of AI projects~\cite{azevedo_kdd_2008,espadinha-cruz_review_2021, forootan_machine_2022,wirth_crisp-dm_2000,xu_review_2022}.
The implementation phases of the methodologies in literature are quite similar, although the naming is different~\cite{azevedo_kdd_2008}.
In this work, the focus is on the phases of the CRISP-DM methodology~\cite{wirth_crisp-dm_2000}.
The main reason for orienting to CRISP-DM is that it is described in the literature as a de-facto standard in the industry and is widely used due to its generality~\cite{saltz_crisp-dm_2020,schroer_systematic_2021}.
Additionally, the phases of CRISP-DM can be applied to other sub-topics of data science projects that are not covered under the term data mining.
CRISP-DM comprises six phases~\cite{wirth_crisp-dm_2000}:

\begin{enumerate}
    \item The \textit{business understanding} phase involves gathering knowledge on the application domain and the project objectives.
    \item The \textit{data understanding} starts with initial data collection and analysis to get familiar with the data.
    \item In the \textit{data preparation} phase, the input datasets are built from the given data by applying techniques such as normalization to transform the data.
    \item The \textit{modeling phase} applies the AI algorithm to the prepared data, and additional fine-tuning is applied, e.g., hyper-parameter optimization.
    \item The \textit{evaluation phase} is used to obtain whether the elaborated model performs as it should.
    \item The \textit{deployment phase} deals with infrastructure and the presentation of customer-usable knowledge.
\end{enumerate}

\subsection{Related Work}\label{sec:background_Related}
Relevant secondary studies started investigating the combination of model-driven software engineering techniques and AI/ML.

%

\luca{added}
\ins{In their studies both Giray~\cite{GIRAY2021} and Martínez-Fernández et al.~\cite{martinez2022} investigate Software Engineering (SE) practices for AI-based systems, or SE4AI.
In particular, \cite{GIRAY2021} highlights the shift from traditional software development to ML systems that learn from data, necessitating reevaluating software development practices. The study systematically reviews 141 studies to analyze the current research on software engineering (SE) for ML systems, revealing that none of the SE aspects have mature tools and techniques, with testing being the most researched area. The paper identifies the need for experiments and case studies in industrial settings to understand challenges and propose solutions, aiming to assist practitioners, researchers, and educators in ML systems engineering. Giray highlights how the non-deterministic nature of ML systems complicates SE, the lack of mature tools, and the importance of practical research to advance the field.}

\ins{In~\cite{martinez2022}, the research highlights the field’s significant expansion since 2015, with a notable focus on dependability and safety. The authors have meticulously mapped prevalent SE methodologies across the areas outlined in the Software Engineering Body of Knowledge (SWEBOK\cite{SWEBOK}).}
\ins{Despite the work’s substantial contribution to bridging the SE and AI disciplines, it points out a notable gap: MDE remains underexplored within SE for AI-based systems. Yet, MDE is acknowledged for its potential to handle complexity and ensure uniformity across various models and artifacts, culminating in more resilient and maintainable software solutions.}

With respect to~\cite{GIRAY2021,martinez2022}, \ins{our research delves deeper into the role of MDE in the context of AI-based systems, referred to as MDE4AI. Specifically, we focus on DSLs utilized within the engineering process and their support in facilitating data mining activities, and we analyze their effectiveness from the perspective of the CRISP-DM framework.}

\ins{To refine the focus from the broader domain of SE4AI to the more specialized field of MDE4AI, we found two surveys that have contributed to this area of research.}

In~\cite{portugal_survey_2016}, Portugal et al. surveyed DSLs and frameworks for designing machine learning algorithms in the context of Big Data. The DSLs were categorized based on the classification proposed by~\cite{czarnecki2005overview,fowler_domain_2010,voelter2013dsl}. However, it is important to note that their survey lacked a systematic approach; no explicit surveying protocol was followed.
In contrast, our approach introduces a classification based on DSL engineering principles and practices commonly used in MDE~\cite{brambilla_model-driven_2017,burgueno_contents_2019}. We emphasize the relationship between DSLs and the implementation phases of AI algorithms, which was not explored by Portugal et al. 
\ugh{Additionally, we acknowledge and utilize the referenced literature from Portugal et al. in our survey for snowballing purposes.}\luca{Did we really do so? It is not mentioned in the protocol, and I would remove such a statement}

\ins{The second survey by Naveed et al.~\cite{naveed2024} is the most interesting and delve into the intersection of MDE and machine learning components, which they term MDE4ML. Their study runs in parallel with our systematic review}\footnote{The preprint of this work has been cited by~\cite{naveed2024}.}, and like ours, it aims to analyze existing research on MDE4ML (a subset of MDEAI). The authors’ review process results in the selection of 46 primary studies, highlighting the increasing adoption of ML as a component in various software applications. This trend is driven by the desire to harness large volumes of data for predictive capabilities and informed decision-making.
Naveed et al.'s study also explores a range of MDE solutions applied to ML systems, including modeling languages, model transformations, and tool support. Notably, the corpus of selected studies partially overlaps with ours~\cite{al-azzoni_model_2020,Baumann2022,bhattacharjee_stratum_2019,giner-miguelez_domain-specific_2022, hartmann_next_2017,koseler_realization_2019,ries_mde_2021}, emphasizing the importance of investigating the integration of MDE and AI/ML in software and system engineering approaches. While both studies provide an overview of MDE techniques and practices within the MDE4AI domain, our systematic literature review (SLR) places particular emphasis on explicitly modeling AI/ML concerns in DSLs and the intent behind related model transformations. Additionally, our analysis of AI/ML support for data mining aligns with the CRISP-DM standard.
In summary, our work provides a complementary viewpoint and results, contributing to the broader understanding of MDE and AI/ML integration in software engineering.

\ins{In~\cite{bucaioni2022} Bucaioni et al. discuss the rise of low-code development (LCD), supporting platforms (LCDP), and its growing interest in the software engineering community, particularly in model-driven engineering (MDE), presenting a multi-vocal systematic review.
Their review highlights the increasing publication trend in low-code research, the prominence of MDE as a core technology in LCD, and the wide range of business domains where LCD is applied. However, the adoption of LCD for developing AI-augmented software systems is still in the early stages. Only one surveyed paper~\cite{Kourouklidis2020} provides insights into the architectural aspects that leverage MDE for monitoring machine learning model performance. However, it does not delve into specific DSLs that could be employed within such a platform.}
 
\ins{The study by Bencomo et al.~\cite{bencomo2019} discusses the principles and requirements of models at runtime and the state of the art.
In this context, machine learning techniques have been reported to support reasoning on uncertainty, building 
context-aware systems, systems-of-systems, and self-aware systems, and extracting information at runtime to dynamically build the models. However, adopting AI/ML is still limited and recognized as a challenge in applied model-driven techniques.} 
\section{Research Methodology}\label{sec:method}
This section introduces the SLR method applied in this work.
The SLR study protocol is based on the guidelines by~\cite{kitchenham_guidelines_2007,kitchenham_systematic_2013,petersen_guidelines_2015}, introducing the main steps of SLRs to be performed in the Software Engineering domain.

\begin{figure*}
    \centering
    \includegraphics[width=1\textwidth]{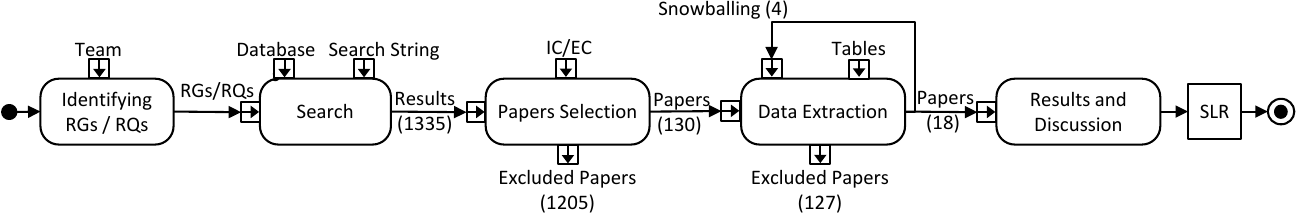}
    \caption{SLR Methodology Overview \luca{do we need to update the numbers in the workflow?}\simon{updated 01.05.2024 (removed arbiter and glinda)}}
    \label{fig:slr-reference-workflow}
\end{figure*}

Figure~\ref{fig:slr-reference-workflow} depicts an activity-like diagram of the performed search and selection process protocol workflow.
The workflow consists of the following steps:

\begin{enumerate}
\item \textit{Identifying the Research Goals and the Research Questions} (RG/RQ): The objective of this work and the research questions are defined to guide the SLR (Section~\ref{sec:introduction} shows the result of the RG/RQ elaboration)
\item \textit{Search Process}: The literature search is conducted on selected databases collecting scientific publications via the execution of queries based on a search string suitably designed according to the given RGs and RQs (Section~\ref{sec:search-process}).
\item \textit{Study Selection}: The authors define the inclusion and exclusion criteria (IC/EC) and apply them to the papers collected in the databases by reading their titles and abstracts. Subsequently, the selected papers are evaluated based on their content (Section~\ref{sec:papers-selection}).
\item \textit{Data Extraction}: 
Detailed information is collected from selected studies that have passed the inclusion and exclusion criteria (IC/EC). This extraction occurs during a thorough full-text reading. The collected data are organized into evaluation tables. Additionally, backward and forward snowballing techniques are applied to selected studies.\luca{rephrased using backward and forward snowballing terms}
\item \textit{Results Analysis and Discussion}: Collected results are analyzed, and a discussion occurs among the authors to answer the stated RQs.
\end{enumerate}

The execution of the protocol is documented in a spreadsheet\footnote{\label{githubfootnote}\url{https://github.com/sraedler/Model-Driven-Engineering4Artificial-Intelligence}}, and bibliographic entries are collected in Zotero Library. An export can be found online\footnoteref{githubfootnote}.

\subsection{Research Questions}\label{sec:RQs}
The overall research goal has been previously introduced in Table~\ref{tab:goal}, aligning with the main research question. To address the main research question, we have defined several refined and more fine-grained research questions as follows \luca{I would remove italics in the explanation of RQs}:

\begin{enumerate}[align=left,label=\textbf{RQ\arabic*},ref=\textbf{RQ\arabic*}]

    \item \label{rq1} What are the prevalent Model-Driven Engineering (MDE) concepts and practices~\cite{brambilla_model-driven_2017,burgueno_contents_2019} being applied in current studies, such as metamodeling and model transformations?
    \item[] \textit{This research question aims to evaluate the application and evidence of MDE concepts and practices.}
    
    \item \label{rq2} Which phases of AI development aligned with the CRISP-DM methodology are covered by the approaches?
    \item[] \textit{This RQ assesses the extent to which the development phases of CRISP-DM are covered. As a result, implications can be made about the extent of support.}
    

    \luca{I rephrased this RQ3 to be more generic since we are not necessarily or only considering industrial domains but any application domain. No EC in this regard, indeed.}
    
    \item \label{rq3} Which application domains actively incorporate Model-Driven Engineering (MDE) methodologies in AI applications?
    \item[] \textit{This RQ aims to identify the specific application domains that actively incorporate MDE methodologies in AI applications. The goal is to understand if any predominant and leading application domain is leading and shaping the evolution of MDE4AI.}
    
    \item \label{rq4} What are the used methods and the supporting MDE tools the proposed approaches rely on?
    \item[] \textit{This RQ allows assessing the underlying methods and the related tool support, including further development leveraging these underlying technologies to gain maturity.}
    
    \item \label{rq5} To what extent is communication between different stakeholders supported by MDE?
    \item[] \textit{Communication and business knowledge elaboration are two of the core pitfalls in the development of AI solutions~\cite{piorkowski_how_2021}. Therefore, this question aims to assess the contribution to support fostering AI in the industry.}
    
    \item \label{rq6} Which challenges and research directions are still open?
    \item[] \textit{This RQ will lead to future research directions and challenges for MDE4AI applications due to a collection of limitations in the proposed approaches based on respective authors or our obtainment.}
    
\end{enumerate}

\subsection{Search Process}\label{sec:search-process}
This section describes the search activity in Figure~\ref{fig:slr-reference-workflow}.
According to~\cite{kitchenham_systematic_2013}, defined search queries are executed on dedicated search engines.
 In this research, the queries are performed on the following bibliographic sources:

\begin{itemize}
  \item ACM Digital Library: \url{http://dl.acm.org/}
  \item dblp Computer Science Bibliography: \url{https://dblp.org/}
  \item IEEE Xplore Digital Library: \url{http://ieeexplore.ieee.org}
  \item Google Scholar: \url{https://scholar.google.com}
  \item Springer: \url{https://link.springer.com}
\end{itemize}

To select suitable terms for the search, keywords from known studies, the MDE4AI\footnote{\url{https://mde-intelligence.github.io/}} \ins{and LowCode}\footnote{\url{https://lowcode-workshop.github.io/}} workshops, and the International Journal on Software and Systems Modeling (SoSyM)\footnote{\url{https://www.sosym.org/}} were selected. 

The selected keywords for the search terms are the following:

\begin{small}
\noindent $S_1(MDE)=\{MDE;Model-Driven~Engineering;Model-Driven\\Development;DSL;Domain~Specific~Language;Metamodeling;Domain~Modeling,\\Low-Code;No-Code;Models@Runtime;Runtime~Models\}$ \\
\noindent $S_2(AI)=\{AI;Artificial~Intelligence;ML;Machine~Learning;DL;\\Deep~Learning;Neural~Network;Data~Science;Intelligence;Data~Analytics\}$
\end{small}

Each keyword $k_{i}$ from the set ${S_1}$ and ${S_2}$ has been combined in conjunctive logic proposition $p\in P$.

\begin{small}
\begin{center}
    \noindent $P = \{p|p=s_i \in S_1 \wedge s_j \in S_2 \}$\\
\end{center}
\begin{center}
    \noindent $\;i=1,2,3,4,5,6,7,8,9,10,11,12\;j=1,2,3,4,5,6,7,8,9,10$\\
\end{center}
\end{small}

The resulting set $P$ of 120 propositions ($p_i$) includes the final \textit{search strings}.
According to~\cite{kitchenham_systematic_2013}, the propositions ($p_i$) should be combined as OR statements.
However, for some search engines, a single search term is too complicated, as some search engines limit the length of the search term or do not generate results correctly due to nested search terms. 
Therefore, each search string is executed as a single query.

The automated search was executed in November 2022 and in February 2024 again.
In total, 1335 papers have been collected.
The search terms and results are archived and are online available\footnoteref{githubfootnote}.
If a result file is unavailable, the search query on the specific search engine did not retrieve any results.

\subsection{Paper Selection}\label{sec:papers-selection}
The inclusion and exclusion criteria (IC/EC) outlined in Table~\ref{tab:ic_ec} are employed for the paper selection.
The IC/EC have been evaluated for each paper collected by queries executed on the selected databases by reading its title and abstract.

\ins{We establish two inclusion criteria for selecting papers. 
The first one is intended to select papers based on an MDE standpoint. We focus on work related to MDE4AI. Our inclusion criteria encompass papers that explicitly present DSLs defining AI/ML extensions as first-class language concepts. These DSLs should allow the specification elements related to AI/ML algorithms like inputs, outputs, parameters, or hyperparameters.}






\ins{We exclude DSLs that merely provide a link to AI/ML artifacts, even if they fall within the field of Model-Driven Engineering for Artificial Intelligence (MDE4AI) (e.g.,~\cite{Atouani2021}).}

\ins{The second one is intended to select papers based on an AI/ML standpoint.
We explicitly exclude any paper adopting AI/ML learning techniques to further enhance MDE approaches (AI4MDE) like modeling recommender based on AI/ML algorithms (e.g.~\cite{morgan}).
We also omit studies focused solely on DSLs that specify algorithms, such as data science workflows (e.g., \cite{delineGlindaSupportingData2021}), unless these DSLs are integrated within a broader AI-augmented software system.}

\del{Although the EC excludes doctoral theses due to their extensiveness, the author's referenced publications are included in snowballing.}

\del{Although review papers are not considered for the survey, we presented relevant surveys in the background section (Section.~\ref{sec:background_Related}).}

\ins{Ultimately, adhering to standard exclusion criteria SLRs, we omitted all secondary studies, non-English publications, vision statements, non-academic articles, proposals, and theses.}

\begin{table}
\centering
\caption{Inclusion Criteria (IC) and Exclusion Criteria (EC)}
\label{tab:ic_ec}
\begin{tabular}{|l|l|l|}
\hline
Type & ID & Type \\ 
\hline
\multirow{4}{*}{IC} & 1 & \begin{tabular}[c]{@{}l@{}}The paper presents a DSL with AI extensions\end{tabular} \\ 
\cline{2-3}
 & 2 & \begin{tabular}[c]{@{}l@{}}The paper leverages MDE body of knowledge, techniques, \\and practices~\cite{burgueno_contents_2019, brambilla_model-driven_2017} \end{tabular}\\ 
\cline{2-3}
 & 3 & \begin{tabular}[c]{@{}l@{}}Full text of the paper is available \end{tabular}\\ 
\hline
\multirow{9}{*}{EC} & \ins{1} & \begin{tabular}[c]{@{}l@{}}\ins{We exclude papers with focus on AI for MDE} \\
 \end{tabular} \\
\cline{2-3}
 & 2 & \begin{tabular}[c]{@{}l@{}} We exclude MDE approaches whose DSLs do not include \\any AI concept as first-class entities\end{tabular}\\ 
\cline{2-3}
\cline{2-3}
 & 3 & We exclude \chg{secondary}{non-primary} studies\\ 
\cline{2-3}
\cline{2-3}
 & 4 & We exclude studies available only in the form of abstract \\ 
\cline{2-3}
 & 5 & We exclude study not in English language \\ 
\cline{2-3}
\cline{2-3}
 & 6 &\begin{tabular}[c]{@{}l@{}} We exclude vision only, non-scientific papers, and proposals\end{tabular} \\
\cline{2-3}
 & 7 & We exclude any thesis \\
\hline
\end{tabular}
\end{table}

Following the IC/EC criteria application, a full-paper read is applied to select the final papers.
Additionally, as suggested by~\cite{kitchenham_systematic_2013}, snowballing is accomplished to retrieve further results.
The relevant papers from the list of forward/backward snowballing papers were selected using the same procedure as the query results.
Table~\ref{tab:final_papers} lists the final list of selected papers aligned with the publication venue.
Particularly, \chg{15}{18} papers are added by query selection~\cite{al-azzoni_model_2020,bergerEVOALDomainSpecificLanguageBased2023,bhattacharjee_stratum_2019,giner-miguelez_domain-specific_2022,hartmann_next_2017,hartmann_meta-modelling_2019,Ming2023,koseler_realization_2019,meacham_adaptivesystems_2021,moin_model-driven_2022,morales_towards_2022,pinedaRADENNDomainSpecificLanguage2023,radlerCodeGenerationMachine2023,ries_mde_2021,schoene2019} and four are added due to snowballing~\cite{de_la_vega_lavoisier_2020,hartsell_model-based_2019,kusmenko_engineering_2019,melchor_model-driven_2022}.

\begin{landscape}
\begin{footnotesize}
\begin{xltabular}{\linewidth}{|p{0.15\textwidth}|X|r|l|X|}

\caption{List of selected publications \del{with type of publication incl. snowballing results}}
\label{tab:final_papers} \\

\hline \multicolumn{1}{|c|}{\textbf{Paper Type}} & \multicolumn{1}{c|}{\textbf{Venue}} & \multicolumn{1}{c|}{\textbf{Year}} & \multicolumn{1}{c|}{\textbf{Author}}  & \multicolumn{1}{c|}{\textbf{Title}} \\ \hline 
\endfirsthead

\multicolumn{3}{c}%
{\tablename\ \thetable{} -- continued from previous page} \\
\hline \multicolumn{1}{|c|}{\textbf{Paper Type}} & \multicolumn{1}{c|}{\textbf{Venue}} & \multicolumn{1}{c|}{\textbf{Year}} & \multicolumn{1}{c|}{\textbf{Author}}  & \multicolumn{1}{c|}{\textbf{Title}} \\ \hline 
\endhead

\hline \multicolumn{5}{|r|}{{Continued on next page}} \\ \hline
\endfoot

\hline
\endlastfoot
\hline
Conference & International Conference on Intelligent Data Science Technologies and Applications (IDSTA) & 2020 & \cite{al-azzoni_model_2020} & Model Driven Approach for Neural Networks \\ 
\hline


Congress & IEEE Congress on Evolutionary Computation (CEC) & 2023 & \cite{bergerEVOALDomainSpecificLanguageBased2023} & EVOAL: A domain-specific language-based approach to optimisation \luca{diff}\\ 
\hline

Conference & IEEE International Conference on Big Data (Big Data) & 2019 &  \cite{bhattacharjee_stratum_2019} & STRATUM: A BigData-as-a-Service for Lifecycle Management of IoT Analytics Applications \\ 
\hline

Journal & Journal of Computer Languages & 2020 & \cite{de_la_vega_lavoisier_2020} & Lavoisier: A DSL for increasing the level of abstraction of data selection and formatting in data mining \luca{diff}\\
\hline


Conference & International Conference on Model Driven Engineering Languages and Systems (MODELS) & 2022 & \cite{giner-miguelez_domain-specific_2022} & A domain-specific language for describing machine learning dataset \\ 
\hline

Journal & \chg{International Conference on Model Driven Engineering Languages and Systems (MODELS)}{Software and Systems Modeling (SoSym)} & \chg{2017}{2019} & \cite{hartmann_next_2017} & The next Evolution of MDE: A Seamless Integration of Machine Learning into Domain Modeling \\ 
\hline

Conference & International Conference on Model Driven Engineering Languages and Systems (MODELS) & 2019 & \cite{hartmann_meta-modelling_2019} & Meta-Modelling Meta-Learning \luca{diff}\\ 
\hline

Workshop & Workshop on Design Automation for CPS and IoT (DESTION) & 2019 & \cite{hartsell_model-based_2019} & Model-based design for CPS with learning-enabled components \luca{diff}\\
\hline

\ins{Journal} & IEEE Transactions on Computer-Aided Design of Integrated Circuits and Systems & 2023 & \cite{Ming2023} & \ins{AIoTML: A Unified Modeling Language for AIoT-Based Cyber–Physical Systems} \luca{diff} \\
\hline

Conference & International Conference on Model-Driven Engineering and Software Development (MODELSWARD) & 2019 & \cite{koseler_realization_2019} & Realization of a Machine Learning Domain Specific Modeling Language: A Baseball Analytics Case Study \\ 
\hline

Conference & International Conference on Automated Software Engineering Workshop (ASEW) & 2019 & \cite{kusmenko_engineering_2019} & On the Engineering of AI-Powered Systems \luca{diff} \\
\hline

Journal & IEEE Access & 2021 & \cite{meacham_adaptivesystems_2021} & AdaptiveSystems: An Integrated Framework for Adaptive Systems Design and Development Using MPS JetBrains Domain-Specific Modeling Environment \luca{diff}\\ 
\hline

Conference & International Conference on Advanced Information Systems Engineering (CAISE) & 2022 & \cite{melchor_model-driven_2022} & A Model-Driven Approach for Systematic Reproducibility and Replicability of Data Science Projects \luca{diff} \\
\hline

Journal & Software and Systems Modeling (SoSym) & 2021 & \cite{moin_model-driven_2022} & A Model-Driven Engineering Approach to Machine Learning and Software Modeling \luca{diff}\\ 
\hline

Journal & International Conference on Product-Focused Software Process Improvement (PROFES) & 2022 & \cite{morales_towards_2022} & Towards a DSL for AI Engineering Process Modeling \luca{diff} \\ 
\hline

Journal & IEEE Access & 2023 & \cite{pinedaRADENNDomainSpecificLanguage2023} & RADENN: A Domain-Specific Language for the Rapid Development of Neural Networks \luca{diff} \\ 
\hline

Conference & Conference Modellierung & 2023 & \cite{radlerCodeGenerationMachine2023} & Code Generation for Machine Learning using Model-Driven Engineering and SysML \luca{diff} \\ 
\hline

Conference & International Conference on Model-Driven Engineering and Software Development (MODELSWARD) & 2021 & \cite{ries_mde_2021} & An MDE Method for Improving Deep Learning Dataset Requirements Engineering using Alloy and UML \\ 
\hline


\hline
\end{xltabular}
\end{footnotesize}
\end{landscape}

\subsection{Data Extraction}\label{sec:data-extraction}
Each selected paper presented in Table~\ref{tab:final_papers} underwent a data extraction process following the data extraction template in Table~\ref{tab:dataextractiontemplate}.
Additionally, the publication type is assessed as Exploratory (without evaluation, e.g., a pure concept or vision) or Technical (with evaluation).
\begin{landscape}
\begin{table}
\centering
\begin{footnotesize}
\caption{Data Extraction Template}
\label{tab:dataextractiontemplate}
\renewcommand{\arraystretch}{1.1}
\begin{tabularx}{\linewidth}{|l|l|l|X|}
\hline
\textit{\textbf{RQ}} & 
\multicolumn{1}{l}
{\textbf{Concern}} & &
\textbf{Assessment Description} \\ 
\hline
\multirow{15}{*}{\textit{RQ1}} & \multirow{15}{*}{MDE} &{\begin{tabular}[c]{@{}l@{}} Language Workbench (LW)\end{tabular}} & A language workbench provides tools to support the definition, reuse, and composition of domain-specific languages (DSLs).\\
 &  & {\begin{tabular}[c]{@{}l@{}}Metamodel(s)/Grammar(s) (MM/G)\end{tabular}} & The metamodel of the approach is either depicted as a diagram in a referenced repository or mentioned and textually described. \\
 &  & Concrete Syntax(es) (CS) & The concrete syntax is given if figures, listings, or tables to illustrate an implementation/use case excerpt or it is indicated whether textual or graphical modeling is applied for a specific aspect. \\
 &  & Validation (V) & The approach takes into account validation concerns. \\
 &  & Model Transformation(s) (MTrafo) & The approach uses or introduces model transformation to generate engineering artifacts of any kind. \\ 
 &  & Development and Runtime Model (D/R) & The approach can use models at development and/or runtime. \\

\hline
\multirow{15}{*}{\begin{tabular}[c]{@{}l@{}}\textit{RQ2,}\\\textit{RQ5}\end{tabular}} & \multirow{15}{*}{AI} & 
 Business Understanding (BU) & The model contributes to the understanding of the underlying business. Particularly, the creation of the data and aspects from other disciplines are introduced, such as requirements modeling for AI.~ \\
 &  & Data Understanding (DU) & The model supports at least two of the following aspects: data description, data attribute interrelation, data background, data quality, and data composition. \\
 &  & Data Ingestion (DI) & The model clearly depicts the origin of data and how to load it. \\
 &  & Feature Preparation (FP) & The model allows an understanding of how data needs to be transformed, connected, or preprocessed. \\
 &  & Model Training (MTrain) & The model depicts the used algorithm with input and output values and potential hyperparameters. \\
 &  & Metrics/Evaluation (ME) & The model depicts metrics for the AI approach or introduces evaluation criteria. \\ 
\hline
\multirow{2}{*}{\textit{RQ3}} & \multirow{4}{*}{Others} & Problem Domain & The domain of the case study or the mentioned area of application. \\ 
\cline{1-1}\cline{3-4}
\multirow{2}{*}{\textit{RQ4}} &  & Frameworks & The method and tools used in the approach, e.g., WebGME, Xtext, Xtend, etc.\\ 
\hline
\end{tabularx}
\end{footnotesize}
\end{table}
\end{landscape}

The extracted data mainly address two concerns of interest, i.e., MDE and AI.
Modeling concerns refer to the evidence of sound knowledge and application of \textit{model foundations}~\cite{burgueno_contents_2019} (e.g., abstract syntax/grammar/metamodel, textual/graphical concrete syntax, constraints, model transformations) and supporting tools (e.g., modeling language frameworks). 
AI concerns~\cite{akkiraju_characterizing_2020} indicate to which extent the publications support ML modeling aligned with the dimensions of the CRISP-DM methodology~\cite{wirth_crisp-dm_2000}. 
It should be noted that the assessment dimensions do not correspond exactly to the phases of CRISP-DM to allow for a more detailed categorization of concerns; e.g., in CRISP-DM, \textit{Data Ingestion} is part of the \textit{Data Understanding} phase but separated in the given assessment.
An aspect of a concern of interest is assessed as \textit{available} ($\blacksquare$) if the aspect is presented in the approach or as \textit{underlying principle} \chg{is}{if it is} typically offered by the underlying environment (e.g., constraint modeling might not be presented but is typically offered by the underlying MDE tooling).
Finally, it is worth noting that there is no evaluation of the deployment phase of CRISP-DM as it is beyond the scope of this paper.

\section{Literature Assessment}\label{sec:results}

\luca{integrate the following statements in the text.}
\todo{OCL/P in MontiCore is a variant of the Object Constraint Language tailored for modeling invariants, method pre/post conditions.}

The result obtained from the data extraction process described in the previous section is presented in Tables~\ref{tab:resultsDataExtraction},~\ref{tab:methodsAndTools} and~\ref{tab:artifacts}.

\begin{landscape}
\begin{table}[]
\centering
\caption{Result of the data extraction for the MDE and AI concerns. \tech{Add again \cite{delineGlindaSupportingData2021} if accepted.}\tech{I want to check whether it will be the only one without a clear language workbench behind. Glinda has a DSL that is basically a set of YAML templates.}}
\vspace{1mm}
\label{tab:resultsDataExtraction}
\resizebox{\linewidth}{!}{%
\begin{tabular}{|c|c|c|c|c|c|c|c|c|c|c|c|c|c|c|c|c|c|c|c|c|c|c|} 
\hline
RQ & Concern & Paper & 
\cite{al-azzoni_model_2020} & \cite{bergerEVOALDomainSpecificLanguageBased2023} & \cite{bhattacharjee_stratum_2019}  &  \cite{de_la_vega_lavoisier_2020} &  
\cite{giner-miguelez_domain-specific_2022} & \cite{hartmann_next_2017} & 
\cite{hartmann_meta-modelling_2019} & 
\cite{hartsell_model-based_2019}  &
\cite{Ming2023} & 
\cite{koseler_realization_2019} & \cite{kusmenko_engineering_2019} & \cite{meacham_adaptivesystems_2021} & 
\cite{melchor_model-driven_2022} & 
\cite{moin_model-driven_2022}  & 
\cite{morales_towards_2022} &  \cite{pinedaRADENNDomainSpecificLanguage2023} &
\cite{radlerCodeGenerationMachine2023} & 
\cite{ries_mde_2021} 
& Sum\\ 
\hline
~ & General & \begin{tabular}[c]{@{}l@{}}\textbf{Type}\end{tabular} & \textbf{T} & \textbf{T} & \textbf{T}  & \textbf{T} & 
\textbf{T} & \textbf{T} & \textbf{E} & \textbf{T}  &\ins{\textbf{T}} 
& \textbf{E} & \textbf{T} & \textbf{T} & \textbf{T} & \textbf{T} & \textbf{E} 
&  \textbf{T} &\textbf{T} 
& \textbf{T} 
& \begin{tabular}[c]{@{}l@{}}15 \textbf{T}\\~3 \textbf{E}\end{tabular} \\ 
\hline
\multirow{10}{*}{RQ1} & \multirow{10}{*}{MDE} & LW & $\blacksquare$ & $\blacksquare$\tech{emf xtext}& $\blacksquare$  & $\blacksquare$ &
$\blacksquare$ & $\blacksquare$ & $\blacksquare$\tech{EMF or Greycat} & $\blacksquare$\tech{GME}  &  $\blacksquare$\tech{EMF xtext}
& $\blacksquare$\tech{papyrus} & $\blacksquare$\tech{monticore} & $\blacksquare$\tech{mps} & $\blacksquare$\tech{emf} & $\blacksquare$\tech{xtext}  & $\blacksquare$\tech{emf} &   & $\blacksquare$\tech{papyrus} & $\blacksquare$\tech{emf} 
& 17\\ 
\cline{3-22}
 &  & MM/G & $\blacksquare$ & $\blacksquare$ &$\blacksquare$  & $\blacksquare$ 
 & $\blacksquare$ & $\blacksquare$ & $\blacksquare$ & $\blacksquare$  &$\blacksquare$ 
& $\blacksquare$ & $\blacksquare$ & $\blacksquare$ & $\blacksquare$ & $\blacksquare$  & $\blacksquare$ &  $\blacksquare$ & $\blacksquare$ & $\blacksquare$ 
& 18\\ 
\cline{3-22}
 &  & CS & xmi & txt & g & txt 
 & txt & txt & txt & g  & txt 
& g & txt/g & txt & xmi & txt  & g & txt  & g & g 
 &\begin{tabular}[c]{@{}l@{}}10 \textbf{txt}\\~7 \textbf{g}\\~2 \textbf{xmi}\end{tabular} \\ 
\cline{3-22}
 &  & V & $\blacksquare$ & $\blacksquare$ & & &
 &  &  &  &
& &  &  & $\blacksquare$ &   & 
& &  
& $\blacksquare$ 
& 4\\ 
\cline{3-22}

 &  & MTrafo & 
 $\blacksquare$ & 
 & 
$\blacksquare$ & 
$\blacksquare$ & 
 $\blacksquare$ & 
 $\blacksquare$ & 
 $\blacksquare$ & 
 $\blacksquare$ & 
 $\blacksquare$ & 
 $\blacksquare$ & 
 $\blacksquare$ & 
 $\blacksquare$ &  
 & 
 $\blacksquare$ & 
 & 
 $\blacksquare$ & 
 $\blacksquare$ & 
 $\blacksquare$ &
 15\\ 

 \cline{3-22}
 &  & D/R & d & d & d & d & d
 & d/r & d & d & d & d & d & d & d & d & d & d & d & d 
 &\begin{tabular}[c]{@{}l@{}}18 \textbf{d}\\~1 \textbf{r}\end{tabular} \\ 

\hline
\multirow{6}{*}{\begin{tabular}[c]{@{}l@{}}RQ2,\\RQ5\end{tabular}} & \multirow{6}{*}{AI} & BU &  & &  &   &&  &  & $\blacksquare$  &$\blacksquare$ 
&  &  &  &  &   &$\blacksquare$ 
&  &$\blacksquare$ 
& $\blacksquare$ 
& 5\\ 
\cline{3-22}
 &  & DU &  &  $\blacksquare$ &  & $\blacksquare$ & $\blacksquare$ & $\blacksquare$ & $\blacksquare$ &   && $\blacksquare$ &  &  & $\blacksquare$ &   & $\blacksquare$ 
&  $\blacksquare$  &$\blacksquare$ 
& $\blacksquare$ 
& 11 \\ 
\cline{3-22}
 &  & DI & $\blacksquare$ &   &$\blacksquare$ & $\blacksquare$ & &  &  & $\blacksquare$  &&  & $\blacksquare$ & $\blacksquare$ & $\blacksquare$ & $\blacksquare$  & $\blacksquare$ 
&  $\blacksquare$ 
&$\blacksquare$ 
&  
& 11\\ 
\cline{3-22}
 &  & FP & $\blacksquare$ &  &$\blacksquare$  &  & &  &  &   &$\blacksquare$ 
&  &  &  & $\blacksquare$ & $\blacksquare$  & $\blacksquare$ 
&  
$\blacksquare$ 
&$\blacksquare$ 
&  
& 8 \\ 
\cline{3-22}
 &  & MTrain & $\blacksquare$ &  &$\blacksquare$  &  & & $\blacksquare$ & $\blacksquare$ & $\blacksquare$  &$\blacksquare$ 
& $\blacksquare$ & $\blacksquare$ & $\blacksquare$ & $\blacksquare$ & $\blacksquare$  & $\blacksquare$ 
&  $\blacksquare$ 
&$\blacksquare$ 
&  
& 14 \\ 
\cline{3-22}
 &  & M/E &  &  &$\blacksquare$  &  & &  &  & $\blacksquare$  &&  & $\blacksquare$ &  &  & $\blacksquare$  & &   
$\blacksquare$ 
&$\blacksquare$ &  
& 6 \\ 
\hline
\end{tabular}
}
\end{table}
\end{landscape}



In~\cite{al-azzoni_model_2020}, Al-Azzoni proposes a model-driven approach to describe ML problems addressed by artificial neural networks.
The approach enables the description of datasets as well as the consuming Multi-Layer Perception (MLP) Neural Networks (NN).
With templates and code generators, executable Java programs can be generated.
The approach is validated using the Pima Indians Diabetes dataset.
Future work of the approach consists of supporting a wider variety of NNs and support for code generation.

In~\cite{bergerEVOALDomainSpecificLanguageBased2023}, Berger et al. present three DSLs allowing to structure requirements for optimization projects relying on evolutionary algorithms. The approach separates and references formalized knowledge of various experts to enable functional and non-functional requirements definition in a textual set of DSLs. However, the integration of hyper-parameter tuning is not given but planned in future.

In \cite{bhattacharjee_stratum_2019}, Bhattacharjee et al. introduce Stratum, a model-driven tool that enables dealing with the lifecycle of intelligent component development.
The platform addresses design-related concerns such as modeling the ML algorithm pipeline, accessing data streams, allocating and properly sizing cloud-based execution platforms, and monitoring the overall system's quality of service.
The primary goal of this work is to support deploying and maintaining various cloud-based execution platforms.
The MDE part of this work is minor and less detailed.


In~\cite{de_la_vega_lavoisier_2020}, De La Vega et al. introduce a DSL that describes datasets to select sufficient data on a high level.
The approach uses a SQL-like textual language to select, combine and filter various data on an attribute level.
The approach aims to increase a dataset's abstraction level to reduce complexity and make using data mining technologies easier.
The outline of future work indicates that the authors want to perform some empirical experiments to assess the approaches usability, learning curve and effectiveness.
Additionally, the extension of the approaches to be applicable in a wider area of application is intended.

\del{In~\cite{delineGlindaSupportingData2021}, DeLine introduces GUI and Language for Interactive Data Analysis (Glinda), a tool that enables to perform live programming using a textual DSL that is immediately executed in a graphical interface. The approach relies on YAML structured commands that are translated to Python code that is executed similarly as if the user performs the interaction on the GUI elements of the live output.}\luca{so far I prefer to remove it}


In~\cite{giner-miguelez_domain-specific_2022}, the DescribeML DSL is proposed to define ML datasets. 
From a DescribeML model, a template with basic information is automatically generated, based on a given dataset.
The provided DSL allows the definition of metadata, data attributes with statistical features and provenance, and social concerns.
This approach aims to improve the understanding of datasets and thus support the replicability of AI projects.
Currently, this work is limited to the dataset description.
Future work aims to describe AI models and other elements of an AI pipeline and integrate with common web browsers.

In~\cite{hartmann_next_2017}, Hartmann et al. present an approach based on so-called micro-learning units at a language definition level.
This work proposes to weave the learning units into domain modeling, due to the high entanglement of learning units and domain knowledge.
For this purpose, the approach allows the definition of DSLs with learned attributes (i.e., what should be learned), how (i.e., algorithm and parameters), and from what (i.e., other attributes and relations).

Hartmann et al. leverage the previous study for meta-learning in~\cite{hartmann_meta-modelling_2019}. This study proposes two generic metamodels for modeling i) ML algorithms and ii) meta-ML algorithms (i.e., algorithms to learn ML ones).
Future work of the authors is to support a wider range of algorithms and parameters.

In~\cite{hartsell_model-based_2019}, a comprehensive modeling environment for learning-enabled components in CPS development is introduced.
The approach supports training, data collection, evaluation, and verification.
It integrates Goal Structuring Notation (GSN) to support assurance and safety cases.
The publication is, among others, part of a research project\footnote{\url{https://modelbasedassurance.org/}} facilitating MDE.
Currently, the approach is only simulation-based. Therefore, future work consists of integrating hardware-in-the-loop to enable verification.

\ins{In \cite{Ming2023}, Ming et al. propose a novel modelling language, AIoTML, to facilitate the development of Cyber-Physical Systems (CPS) that leverage the power of Artificial Intelligence of Things (IoT). The proposed AIoTML DSL extends ThingML, preserving compatibility and introducing new language constructs to define new datatypes and AI strategies based on deep learning and reinforcement learning models and mechanisms for physical modelling and simulation, enabling the construction of digital twins and optimization of control strategies for CPSs.}


In~\cite{koseler_realization_2019}, a DSL is introduced with the goal of proving the plausibility of using MDE approaches to create ML software.
The DSL, conceptually sketched by another research group in~\cite{breuker_towards_2014}, is realized and applied to a case study in the sports domain. The approach integrates model transformation to generate executable code.
Future work consists of extending the approach to other use cases and conduct empirical studies.

In~\cite{kusmenko_engineering_2019}, an approach describing deep learning using MDE is presented.
The approach combines two DSLs, namely MontiAnna,  and EmbeddedMontiArc. The former is a textual modeling framework for designing and training Artificial Neural Networks (ANNs). It also embeds another DSL, MontiAnnaTrain, for describing the training procedure.
The latter, EmbeddedMontiArc,  is an architectural description language. It supports the definition of components and connectors, with a particular focus on embedded, automotive, and cyber-physical systems.
The frameworks are intended to define deep artificial neural networks, e.g., convolutional neural networks, for processing traffic images to learn how to drive a car in a simulator. 
At this point, it should be noted that only the journal publication by Kusmenko et al. was taken into account, although there are several comparable or extension approaches in the literature list of the author and the associated research group\footnote{\url{https://scholar.google.de/citations?hl=de&user=bUjXPaUAAAAJ}}. The reason for this is that the MDE dimension is overall the same and only small changes in the AI assessment, such as support for different algorithms, are shown.

In~\cite{meacham_adaptivesystems_2021}, Meacham et al. propose a set of DSLs and toolset implemented on top of the MPS\footnote{Meta Programming System} language workbench for the design and development of adaptive
systems offering MAPE-K and AI in context capabilities. The approach describes an extension and composition of DLSs that are extended with application-specific concepts.
Future work consists of extending the range of domains the approach is applicable.

In~\cite{melchor_model-driven_2022}, Melchor et al. propose an MDE approach to formalizing ML projects and the associated infrastructure in which the resulting tool will be deployed.
The approach aims to increase the reproducibility and replicability of data science projects. Hence a key feature of the approach is to describe processes and datasets in detail. With respect to this, future work aims to add verification on a model level to support this core competence.

In~\cite{moin_model-driven_2022}, Moin et al. present an MDE approach based on ThingML\footnote{\url{https://github.com/TelluIoT/ThingML}} to support the development of IoT devices with the extension of data analytics and ML.
The ThingML framework supports defining software parts and components using UML.
The communication between the components (things) is defined using ports, messages, and state machines. The approach supports the transformation of the model into executable code.
Future work aims to extend the approach to more methods, to other technologies such as semi-supervised learning and to enable the generation of various other target languages.

In~\cite{morales_towards_2022}, Morales et al. provide a DSL to model AI-related processes using Eclipse-based technologies.
The approach aims to describe AI processes within an organization and thus contribute to the structured designing, enacting, and automating of AI engineering processes. 
Future work consists of extending the approach to be connectable e.g., with the domain needs. 

In~\cite{pinedaRADENNDomainSpecificLanguage2023}. Pineda et al. present an imperative DSL calls RADENN, allowing to model deep learning using a textual syntax.
The approach defines a grammar constituted by 27 rules with various built-in functions such as \textit{general} functions like 'print' or 'save', \textit{dataset} functions enabling to load and store datasets and \textit{network} functions related to neural network functions like train, predict or evaluate.
The approach aims to enable deep learning modeling with a focus on simplified syntax and increased efficiency while maintaining the same results.
In addition, the approach offers the possibility of practicing online learning, a technique in which training can be paused and continued with another data set.
Future work consists of extending the approach with other NNs.

In~\cite{radlerCodeGenerationMachine2023}, 
Raedler et al. present a SysML-based approach allowing to graphically model machine learning algorithms. The approach is based on hierarchical organized stereotypes that are used to detail either functions or artifacts. Due to the integration in SysML, the approach is well-aligned with CRISP-DM. Based on the stereotypes and the mapping of stereotypes to implementation templates, the approach leverages on model transformation to streamline the implementation of machine learning algorithms~\cite{radlerCodeGenerationMachine2023}.
Future work includes the systematic backflow of information and the extension of the approach to be more readily applicable.

An MDE approach for defining dataset requirements is introduced in~\cite{ries_mde_2021}.
It focuses on the structural definition of requirements using semi-formal modeling techniques.
Future work of the approach aims in rigorous validation of the models and adaption by model-to-model transformation improvement.

\del{In~\cite{zucker_arbiter_2020}, Zucker et al. present a very preliminary version of a declarative DSL for ethical AI addressing transparency, fairness, accountability, and reproducibility concerns of ethical machine-learning datasets.
The approach describes datasets in a SQL-fashioned language and provides a notation to record how ML models will be trained.}\luca{low-quality and too preliminary paper}

\subsection{Model-Driven Engineering Concerns}\label{sec:resultMDE}
In this section, we report the contributions of the selected studies with respect to MDE techniques and practices~\cite{brambilla_model-driven_2017,burgueno_contents_2019}.

In this section, we analyze the adoption of MDE techniques and practices by the selected studies on several key aspects:
\begin{itemize}
    \item Language Workbench \textit{(LW)} Adoption: We explore whether LWs have been utilized for metamodel \textit{(MM)} or grammar \textit{(G)} specifications. Specifically, we investigate whether textual \textit{(txt)} and/or graphical \textit{(g)} concrete syntax \textit{(CS)} is available within these workbenches.
    \item Validation \textit{(V)}: We assess whether the studies consider validation routines via model constraint definitions or model transformations to other engineering artifacts.
    \item Model Transformations \textit{(MTrafo)}: We investigate whether the proposed MDE approaches include model transformations
    \item Applicability at Development \textit{(d)} and/or Runtime \textit{(r)}: We examine whether the proposed approach is applicable during development (design and implementation) and/or runtime (execution). 
\end{itemize}

The analysis results are presented in Table~\ref{tab:resultsDataExtraction}.


\subsubsection*{\textbf{Language Workbenches, Metamodels, and Grammars}}

With the term language workbench, we refer to a tool or set of tools designed for software development within the language-oriented programming paradigm. It typically includes tools to support the definition, reuse, and composition of DSLs through a dedicated integrated development environment~\cite{iung_systematic_2020}.

All the selected studies present metamodels, grammars, or UML profiles to define the concepts of the proposed DSLs.


\todo{Hidden text here, correct but removed because redundant. Uncomment and check if needed.}


In particular, the proposed DSLs have been developed using the following LWs.

\paragraph{EMF-based}
\ins{In this set, we consider approaches directly relying on the EMF LW. Therefore, in addition to approaches proposing DSLs based on Ecore-based metamodels, it includes EMF-based appraches like Xtext-based grammars, and UML profiles.}
In 10 studies, the metamodel is based on EMF~\cite{al-azzoni_model_2020,bergerEVOALDomainSpecificLanguageBased2023,de_la_vega_lavoisier_2020,Ming2023,koseler_realization_2019,melchor_model-driven_2022,moin_model-driven_2022,morales_towards_2022,radlerCodeGenerationMachine2023,ries_mde_2021}\hiddencite{added \cite{Ming2023,koseler_realization_2019,radlerCodeGenerationMachine2023}, removed \cite{jahicSEMKISDSLDomainSpecificLanguage2023}}.
Several EMF metamodels focus on the description of datasets~\cite{al-azzoni_model_2020,de_la_vega_lavoisier_2020,ries_mde_2021}\hiddencite{removed \cite{jahicSEMKISDSLDomainSpecificLanguage2023,giner-miguelez_domain-specific_2022}}.
Other studies additionally describe algorithms~\cite{al-azzoni_model_2020,koseler_realization_2019,melchor_model-driven_2022} or even further steps of the implementation~\cite{Ming2023,morales_towards_2022}\hiddencite{added \cite{Ming2023}}.
\ins{~\cite{bergerEVOALDomainSpecificLanguageBased2023}, three Xtext-based DSLs for data description, input and output data definitions for machine-learning processes, and to configure optimization algorithms.}
In \cite{koseler_realization_2019}, the conceptual metamodel presented as an entity-relationship diagram in ~\cite{breuker_towards_2014} is realized as a UML profile, i.e., a lightweight extension of the EMF-based UML metamodel in Papyrus.
In~\cite{radlerCodeGenerationMachine2023}\hiddencite{removed \cite{radlerModelDrivenEngineeringMethod2023}}, the metamodel is described using SysML stereotypes, an extension of Papyrus UML and, therefore, leverages lightweight extension of UML metamodel. Both~\cite{Ming2023,moin_model-driven_2022} extends an existing Xtext-based DSL, ThingML~\cite{thingml}, with language constructs for AI/ML concerns.

\paragraph{KMF/Greycat-based}
Two studies~\cite{hartmann_next_2017,hartmann_meta-modelling_2019} of the same research group are based on the Kevoree Modeling Framework (KMF) and its successor GreyCat, which results from a research project to create an alternative to the EMF based on Ecore.
In~\cite{hartmann_next_2017}, the capabilities of the Greycat metalanguage are presented. In particular, it allows the definition of microlearning units by explicitly declaring \textit{learned attributes} as part of the domain-specific metamodels.
\ins{In~\cite{hartmann_meta-modelling_2019}, 
Hartmann et al. introduce an exploratory approach proposing two metamodels for ML and meta-learning without specifically selecting a target LW. This work lack an implementation. However, the authors explicitly suggest that both the Greycat and EMF LWs are viable candidates for implementing their approach.}

\paragraph{WebGME-based} 
Two studies~\cite{bhattacharjee_stratum_2019,hartsell_model-based_2019} define metamodels using the WebGME language workbench.
While UML and profiles cannot provide the language engineering support typically offered by language workbenches, WebGME allows specifying DSLs, creating a class diagram-based metamodel from which the DSL infrastructure is automatically generated.
In~\cite{bhattacharjee_stratum_2019}, the so-called Stratum approach for BigData-as-a-Service provides a DSML consisting of several metamodels built on top of WebGME (metamodel for ML algorithms, metamodel for data ingestion frameworks, metamodel for data analytics applications, metamodels for heterogeneous resources).
In~\cite{hartsell_model-based_2019}, the metamodel is based on existing metamodel libraries: SEAM, DeepForge, and ROSMOD.

\paragraph{Langium-based} In \cite{giner-miguelez_domain-specific_2022}, the DescribeML DSL is the only work leveraging the recent Langium open-source language workbench 
enabling domain-specific languages
in VS Code, Eclipse Theia, and web applications, leveraging the Language Server Protocol (LSP)\footnote{\url{https://microsoft.github.io/language-server-protocol/specifications/lsp/3.17/specification/}}.
In~\cite{giner-miguelez_domain-specific_2022}, three metamodels are described i)~metadata model, ii)~composition model, and iii)~provenance and social concerns model. Such metamodels are then implemented as grammars\footnote{Based on Chevrotain, \url{https://chevrotain.io/docs/}.}.

\paragraph{MontiCore-based}
In~\cite{kusmenko_engineering_2019}, all DSLs, i.e., MontiAnna, MontiAnnaTrain, and EmbeddedMontiArc, are all defined using the MontiCore language workbench~\cite{rumpe_monticore_2017}. One of the main benefit is the reuse of existing C++ code generators for neural network frameworks (MxNet, Caffe2, and Tensorflow).

\paragraph{MPS-based} In~\cite{meacham_adaptivesystems_2021}, five different DSLs are created with JetBrains MPS, an open-source projectional language workbench that allows direct changes to the abstract syntax tree through an editor, without the need for a grammar or parser. \cite{meacham_adaptivesystems_2021} leverages MPS' language extension and composition capabilities to deal with domain-independent (e.g., using the AdaptiveSystems DSL to structure the system according to MAPE-K loop by IBM) and domain-specific concerns (e.g., AdaptiveVLE to model concerns of virtual learning environments).

\paragraph{EBNF-based} In~\cite{pinedaRADENNDomainSpecificLanguage2023}, the grammar of RADENN, a domain-specific language for neural networks, is defined using the Extended Backus-Naur Form (EBNF). 
The suite of tools for RADENN, developed in Python, facilitates the language’s application. However, the methodology behind the toolset’s creation—whether handcrafted or auto-generated via a Python-centric language workbench—remains unspecified (e.g.,~\cite{dejanovic2017textx}).

\subsubsection*{\textbf{Concrete Syntax}} 
This section assesses the proposed approaches' notations or \textit{concrete syntax}.

Ten studies
\cite{bergerEVOALDomainSpecificLanguageBased2023,de_la_vega_lavoisier_2020,giner-miguelez_domain-specific_2022,hartmann_next_2017,hartmann_meta-modelling_2019,Ming2023,kusmenko_engineering_2019,meacham_adaptivesystems_2021,moin_model-driven_2022,pinedaRADENNDomainSpecificLanguage2023} 
provide a textual (or tabular) notation;
six studies
\cite{bhattacharjee_stratum_2019,hartsell_model-based_2019,koseler_realization_2019,kusmenko_engineering_2019,morales_towards_2022,radlerCodeGenerationMachine2023,ries_mde_2021} 
\hiddencite{removed \cite{jahicSEMKISDSLDomainSpecificLanguage2023,radlerModelDrivenEngineeringMethod2023}} 
adopt a graphical notation. 


\paragraph{No concrete syntax available}
Two studies~\cite{al-azzoni_model_2020,melchor_model-driven_2022} do not provide a DSL-specific concrete notation. 
In particular, Al-azzoni~\cite{al-azzoni_model_2020} left the definition of a complete DSL as future work while~\cite{melchor_model-driven_2022} is conceived to reuse the notations offered by tools defining data science pipelines. However, by leveraging EMF, a tree-based notation is possible by automatically generated editors, and, potentially, compatible technologies can provide textual or graphical concrete syntax options (e.g., via Xtext and Sirius, respectively). In Table~\ref{tab:resultsDataExtraction}, we reported the serialization format XMI as \textit{default} notation since it can be used by users to inspect the model artifacts.


\paragraph{Textual notation}

In~\cite{bergerEVOALDomainSpecificLanguageBased2023,de_la_vega_lavoisier_2020,Ming2023,moin_model-driven_2022}, all DSLs are defined in Xtext. \cite{Ming2023,moin_model-driven_2022} extends the ThingML grammar, thus the generated artifacts are declared to be still compatible with the original Xtext-based textual editors.
In~\cite{giner-miguelez_domain-specific_2022}, the textual concrete syntax is defined by a recent language workbench, Langium.
In~\cite{hartmann_next_2017} and~\cite{hartmann_meta-modelling_2019}, an Emfatic-inspired textual modeling language.
In~\cite{meacham_adaptivesystems_2021}, five different interwoven DSLs, 
are proposed, mixing textual and tabular projections, created with JetBrains MPS.
\del{In~\cite{zucker_arbiter_2020}, a SQL-like textual notation is proposed. However, they do not provide any grammar, and then the textual notation is just a proposal.}
In~\cite{pinedaRADENNDomainSpecificLanguage2023}, Python-inspired textual notation is applied to model neural network concerns.
\del{In~\cite{delineGlindaSupportingData2021}, a textual, YAML-based Visual Studio extension is described as input for data science operations.}

\paragraph{Graphical notation}
In~\cite{bhattacharjee_stratum_2019} and~\cite{hartsell_model-based_2019}, the graphical concrete syntax is defined through capabilities offered by the WebGME language framework.
\cite{koseler_realization_2019} implements the metamodel as a UML profile in Papyrus. The UML Class Diagram is chosen as graphical notation since all the stereotypes inherit from the Class metaclass. No DSL-specific customization of the UML graphical notation is offered.
\cite{morales_towards_2022} provides a web-based graphical editor realized using Sirius Web\footnote{\url{https://www.eclipse.org/sirius/sirius-web.html}}. In~\cite{ries_mde_2021}\hiddencite{removed \cite{jahicSEMKISDSLDomainSpecificLanguage2023}}, the DSL provides a graphical concrete syntax and editor realized in Sirius\footnote{\url{https://www.eclipse.org/sirius}}. However, the paper does not discuss or show its graphical elements.
In~\cite{radlerCodeGenerationMachine2023}\hiddencite{removed \cite{radlerModelDrivenEngineeringMethod2023}}, graphical modelling using SysML and Papyrus is presented.

\paragraph{Multiple notations}


In~\cite{kusmenko_engineering_2019}, Kushmenko et al. are the only ones proposing a mix of textual and graphical concrete notations to represent AI concerns. However, it is worth noting that the SVG-based hierarchical representation of components and connectors is made for visualization purposes and is not editable\footnote{\url{https://github.com/EmbeddedMontiArc/Documentation}}.



\subsubsection*{\textbf{Validation}}

\luca{new paragraph}

Validation plays a pivotal role in MDE approaches. Modeling constraints assist modelers in creating valid models, which are essential for successfully executing automated engineering processes. While language constraints are typically enforced by metamodels and grammars, modelers can also define additional arbitrary constraints to perform more complex validation tasks on model artifacts. More sophisticated validation techniques, such as formal checks, may involve different artifacts, which are often woven together by model transformations.

In our research, we identified four studies that mention these constraints~\cite{al-azzoni_model_2020,bergerEVOALDomainSpecificLanguageBased2023,melchor_model-driven_2022,ries_mde_2021}.

Leveraging LW can yield a clear advantage in modeling and validation. LW can provide standards for constraint definitions languages like OCL and leverage ad-hoc extensions of engineering platforms. 


For example, the approach presented by Al Azzoni~\cite{al-azzoni_model_2020} leverage the Epsilon Validation Language (EVL) by the Epsilon to ensure the validity of learning problem and neural network models.

In~\cite{bergerEVOALDomainSpecificLanguageBased2023}, Berger et al. provide two separate Xtext-based data definition and constraint definition DSLs for textual specification of single and combined constraints on data, respectively.

In \cite{melchor_model-driven_2022}, constraints are required for satisfying reproducibility and replicability requirements of data science projects' pipelines. OCL constraints are mentioned even though not shown. Their approach leverage EMF
that provides OCL support for EMF-based metamodels (e.g., OCLinEcore~\footnote{\url{https://wiki.eclipse.org/OCL/OCLinEcore}}).

Finally, in~\cite{ries_mde_2021}, the Unified Modeling Language (UML) is employed for semi-formal modeling. To validate formal constraints, the resulting models must be translated into Alloy for formal analysis. The Alloy language facilitates the validation of structural requirements for datasets in Deep Learning-based systems.

\subsubsection*{\textbf{Model Transformation}}
Fifthteen selected studies explicitly mention or report model transformations as part of the proposed approaches.
These model transformations are classified based on their intents, as described in~\cite{lucio_model_2016}, and the technology they use, as described in~\cite{kahani_survey_2019}. Table \ref{tab:ModelTransformationIntent}  summarizes the intents of the model transformation for each paper, as well as the main model-driven technologies used.
It is important to note that none of the papers explicitly list or classify their model transformations. The identification of existing transformations and their intents is an attempt by the authors of this paper to provide a basis for comparison.

\begin{table}[]
\caption{Model transformation intent category and concrete intent\todo{\cite{delineGlindaSupportingData2021} is hidden}}
\label{tab:ModelTransformationIntent}
\centering
\resizebox{\linewidth}{!}{%
\begin{tabular}{|l|l|l|l|} 
\hline
Paper & \textbf{Intent Category} & \textbf{Concrete Intent} & \textbf{Tool} \\ 
\hline
\cite{al-azzoni_model_2020} & Refinement & Synthesis (model to code) & Epsilon EGL\\ 
\hline
\cite{bhattacharjee_stratum_2019} & Refinement & Synthesis (model to code) & JS Implementation\\ 
\hline
\multirow{2}{*}{\cite{de_la_vega_lavoisier_2020}} & Language Translation & Translation & \multirow{2}{*}{Xtend}\\ 
\cline{2-3}
 & Abstraction & Restrictive Query &\\
 \hline
\cite{giner-miguelez_domain-specific_2022} & Language Translation & Translation & \begin{tabular}[c]{@{}l@{}}Typescript \\ (Visual Studio Code)\end{tabular}\\ 
\hline
\multirow{2}{*}{\cite{hartmann_next_2017}} & Refinement & Synthesis (model to code) & \multirow{2}{*}{KMF/GreyCat}\\ 
\cline{2-3}
 & Semantic Definition & Translational Semantics &\\ 
\hline
\multirow{2}{*}{\cite{hartmann_meta-modelling_2019}} & Refinement & Refinement & \multirow{2}{*}{n.a.}\\ 
\cline{2-3}
 & Refinement & Synthesis (model to code) &\\ 
\hline
\multirow{3}{*}{\cite{hartsell_model-based_2019}} & Refinement & Synthesis (model to code) & \multirow{3}{*}{n.a.} \\ 
\cline{2-3}
 & Semantic Definition & Translational Semantics &\\ 
\cline{2-3}
 & Analysis & Safety  Analysis &\\ 
\hline
\multirow{1}{*}{\ins{\cite{Ming2023}}} & \ins{Refinement} & \ins{Synthesis (model to code)} & \begin{tabular}[c]{@{}l@{}} compilers written in Java \end{tabular}\\ 
\cline{2-4}
\hline

\cite{koseler_realization_2019} & Refinement & Synthesis (model to code) & Epsilon EGX/EGL/EOL\\ 
\hline
\cite{kusmenko_engineering_2019} & Refinement & Synthesis (model to code) & \begin{tabular}[c]{@{}l@{}}EmbeddeMontiArc\\/EMADL2CPP\end{tabular}\\ 
\hline
\multirow{2}{*}{\cite{meacham_adaptivesystems_2021}} & Refinement & Synthesis (model to code) & \multirow{2}{*}{Jetbrains MPS}\\ 
\cline{2-3}
 & Model Composition & Model Merging &\\ 
\hline
\cite{moin_model-driven_2022} & Refinement & Synthesis (model to code) & Xtend \\
\hline
 \cite{pinedaRADENNDomainSpecificLanguage2023}& Language Translation & Translation & Python \\
 \hline
 \cite {radlerCodeGenerationMachine2023}& Refinement & Synthesis (model to code) & Python\\ 
\hline
\cite{ries_mde_2021} & Semantic Definition & Translational Semantics  & Xtend\\
\hline
\end{tabular}
}%
\end{table}

\ugh{Eleven} studies leverage model-to-code transformations~\cite{al-azzoni_model_2020,bhattacharjee_stratum_2019,hartmann_next_2017,hartmann_meta-modelling_2019,hartsell_model-based_2019,Ming2023,koseler_realization_2019,meacham_adaptivesystems_2021,moin_model-driven_2022,radlerCodeGenerationMachine2023}\hiddencite{added \cite{Ming2023}, removed \cite{radlerModelDrivenEngineeringMethod2023}} to perform refinements on involved artifacts to generate executable code.
Three studies~\cite{hartmann_next_2017,hartsell_model-based_2019, ries_mde_2021}\hiddencite{removed \cite{jahicSEMKISDSLDomainSpecificLanguage2023}} aim at executable models by defining translational semantics for their DSLs.
\ugh{Seven} approaches \cite{de_la_vega_lavoisier_2020,hartmann_next_2017,hartmann_meta-modelling_2019,hartsell_model-based_2019,Ming2023,pinedaRADENNDomainSpecificLanguage2023}\hiddencite{added \cite{Ming2023} removed \cite{delineGlindaSupportingData2021}} provide more than one transformation with different intents.
Two approaches~\cite{de_la_vega_lavoisier_2020, giner-miguelez_domain-specific_2022} translate artifacts across different modeling languages.

The rightmost column in Table \ref{tab:ModelTransformationIntent} mentions the main model-driven technology leveraged by the studies to implement model transformations.
The most commonly used platform among the studies is Eclipse, with Epsilon\footnote{\url{https://www.eclipse.org/epsilon}} and Xtend\footnote{\url{https://www.eclipse.org/xtend}} being the most popular tools. 

For example, in~\cite{al-azzoni_model_2020}, the Epsilon Generation Language (EGL) is used in conjunction with templates to define model transformations that generate Java code. 
Similarly,~\cite{koseler_realization_2019} uses EGL to generate C\# code for making predictions on test data. 
In~\cite{bhattacharjee_stratum_2019}, WebGME’s code generation capabilities are extended with templates for each sub-task. 
In~\cite{de_la_vega_lavoisier_2020}, two intents of model transformations are reflected: language translation and abstraction using a restrictive query. Based on Xtend, the model transformation transforms dataset descriptions into tabular datasets using low-level data transformation operations, which can then be used in data mining algorithms. 
\del{In~\cite{delineGlindaSupportingData2021}, a compiler transforms YAML code into executable Python commands.}
In~\cite{hartmann_next_2017}, the GreyCat framework, built on the KMF, provides code generation toolsets for building object-oriented applications. 
In~\cite{hartmann_meta-modelling_2019}, the concept of using code generators to generate ML code is mentioned. 
In~\cite{hartsell_model-based_2019}, the ALC toolchain enables code generation for data collection or training exercises of learning-enabled components, as well as translational semantics for configuring an embedded Jupyter Notebook that executes the learning model. The approach also allows for the construction of safety cases. 
\ins{In \cite{Ming2023}, model-to-code transformations are implemented in Java within compilers that bind platform-independent AIoTML models to specific IoT platforms}. 
In~\cite{kusmenko_engineering_2019}, the MontiAnna2X code generator generates MxNet, Caffe2, or Tensorflow code. 
In~\cite{meacham_adaptivesystems_2021}, JetBrains MPS language is used to generate Java code. 
In~\cite{moin_model-driven_2022}, Java and Xtend are used to generate Python code.  
In~\cite{ries_mde_2021}\hiddencite{removed \cite{jahicSEMKISDSLDomainSpecificLanguage2023}}, model-to-code transformation is used to complete formal specifications using the Alloy Analyzer. 
In~\cite{pinedaRADENNDomainSpecificLanguage2023}, an interpreted language is presented. The approach first transforms the code into tokens and trees, further evaluating the syntax and interpreting the defined model tree.
In~\cite{radlerCodeGenerationMachine2023}\hiddencite{removed \cite{radlerModelDrivenEngineeringMethod2023}}, means of model transformation is used to generate executable code based on templates that are mapped with properties of the SysML model enriched with unambiguous stereotypes.

\subsubsection*{\textbf{Development and Runtime Model}}

With term runtime model, \cite{bencomo2019} Bencomo et al. refer to a concept that involves the use of models during the runtime of a system to provide intelligent support and enable self-adaptation.

All the selected studies in this SLR propose model-driven approaches designed for use during various development stages.

Among the selected studies, Hartmann et al.~\cite{hartmann_next_2017}  are the only that explicitly leverage runtime models. In their paper, they propose a design model that is instantiated at runtime, treated as an evolving object graph, and retains historical data. They introduce the concept of microlearning units, which are specified at development time within domain classes that lack a predefined behavioral model. These microlearning units are designed to learn and adapt at runtime based on the data associated with the object graph, effectively enabling the runtime model to acquire \textit{known unknowns}, i.e., behaviors that were not predictable at design time but can be learned through live data.

\subsection{Artificial Intelligence Concerns}
\label{sec:resultAI}
Same as for the MDE concerns, the findings regarding AI development characteristics are presented in the following.

\subsubsection*{\textbf{Business Understanding}}
Industry often faces the problem of missing business understanding and shortcomings in elaborating business values~\cite{brunnbauer_idea-ai_2021,brunnbauer_top-down_2022,radler_survey_2022,someh_overcoming_2020}.
Therefore, modeling business understanding is essential for mature and comprehensive approaches, e.g., by defining requirements.
The assessment revealed that \chg{four}{five} of the 18 approaches foster business understanding by integrating system-relevant modeling or processes.

In~\cite{hartsell_model-based_2019}, the business understanding is fostered due to requirements and components modeling using SysML.
Particularly, a Goal Structuring Notation (GSN) approach is used to define and structure requirements.

In~\cite{morales_towards_2022}, business-relevant information is modeled through integrating \textit{Roles}, leading to increased business understanding.
Additionally, the metamodel reflects means to model requirements.
However, details are currently missing on how the modeling is carried out.

\ins{In~\cite{Ming2023}, the business understanding is defined by so-called Things, which are constructs that are standard in the underlying concept of the modelling language ThingML. Particularly, devices, controllers and simulators are defined as Things.}


In~\cite{ries_mde_2021}\hiddencite{removed \cite{jahicSEMKISDSLDomainSpecificLanguage2023}}, a method to describe ML datasets from a requirements engineering perspective is presented.
Notably, functional and non-functional requirements are integrated to describe dataset structural requirements.




In~\cite{radlerCodeGenerationMachine2023}\hiddencite{removed \cite{radlerModelDrivenEngineeringMethod2023}}, business understanding is explicitly given by means of systems modelling, enabling to describe the origin of the data collected and utilized by the machine learning approach.

\subsubsection*{\textbf{Data Understanding}}
The data understanding fosters the downstream processes of CRISP-DM.
Additionally, it allows assessing dataset quality and streamlining to form hypotheses for hidden
information~\cite{wirth_crisp-dm_2000}.
In the selected literature, 11 approaches support modeling some aspects of the data understanding.


\cite{de_la_vega_lavoisier_2020} contextualizes dataset properties and improves data understanding by implicitly applying rules on how to select data.
In~\cite{giner-miguelez_domain-specific_2022}, a detailed description of a dataset and data composition is given that fosters the overall data understanding.
In~\cite{koseler_realization_2019}, data understanding is enhanced  due to the input data's graphical representation and the variables' composition.
In~\cite{melchor_model-driven_2022}, data understanding is promoted by describing data attributes such as the data type. 
Furthermore, the type of ML algorithm is described, allowing the reproduction of an ML project.
In~\cite{morales_towards_2022}, data understanding is promoted by defining data attributes aligned with attribute types using UML class diagram.

In~\cite{hartmann_next_2017,hartmann_meta-modelling_2019}, the enrichment of properties on a metamodel level is enabled, contributing to further description of the properties and, therefore, increasing data understanding. 
Moreover, the interconnection of the data properties is highlighted by the underlying principle.
Still, the description of the attributes is not very detailed, leading to no support in understanding a single property and its origin.
In~\cite{ries_mde_2021}\hiddencite{removed \cite{jahicSEMKISDSLDomainSpecificLanguage2023}}, the advanced requirements modeling allows a better understanding of datasets with specific properties and structured data elements.

In~\cite{pinedaRADENNDomainSpecificLanguage2023}, the data understanding can be modelled due to the build-in functions and the automatic type inference, which implicitly cast variables based on the use.

In~\cite{bergerEVOALDomainSpecificLanguageBased2023}, data understanding is the primary goal of the approach with separation of concerns of the various information suppliers.

In~\cite{radlerCodeGenerationMachine2023}\hiddencite{removed \cite{radlerModelDrivenEngineeringMethod2023}}, data understanding is given as each property of the provided data is formalized and data types are specified using stereotypes.


\subsubsection*{\textbf{Data Ingestion}}
11 of the given 18 approaches describe the loading and ingestion of data, i.e., loading or referencing the input datasets.

In~\cite{meacham_adaptivesystems_2021}, the implementation of data ingestion using a DSL is described.
Ten other approaches support the specification of a file path, URI, URL, etc., to reference data~\cite{al-azzoni_model_2020,de_la_vega_lavoisier_2020,hartsell_model-based_2019,melchor_model-driven_2022,morales_towards_2022,pinedaRADENNDomainSpecificLanguage2023,radlerCodeGenerationMachine2023}\hiddencite{removed \cite{radlerModelDrivenEngineeringMethod2023}}.
In~\cite{de_la_vega_lavoisier_2020}, the loading of the dataset is described by specifying the name and path of the file or SQL server in combination with SQL selection scripts.
Therefore, this approach supports both file and database-related data.
In~\cite{morales_towards_2022}, data loading from various sources, such as SQL servers, is supported.

In contrast, to fix data sources, the loading from edge devices or sensors is supported by three approaches~\cite{bhattacharjee_stratum_2019,kusmenko_engineering_2019,moin_model-driven_2022}.
In~\cite{bhattacharjee_stratum_2019}, data loading from various edge devices is presented using technologies such as RabittMQ or Kafka. 
In~\cite{kusmenko_engineering_2019}, data loading is provided with tagging schemas for EMADL ports.
In~\cite{moin_model-driven_2022}, two approaches are given, first a black-box approach, where the ML model is imported from a pickle, and second, the paths or URLs of the dataset(s) are passed to the training, validation, and testing of the algorithm.




\subsubsection*{\textbf{Feature Preparation}}
The preparation of features for certain ML algorithms is supported by eight of the 18 approaches.

In~\cite{bhattacharjee_stratum_2019}, the feature preparation is defined in the metamodel. 
Unfortunately, details on the specific methods, parameters, or the order of execution are missing.
In~\cite{al-azzoni_model_2020}, normalization of dataset features is supported. 
However, other pre-processing methods are not supported in the metamodel.
In~\cite{melchor_model-driven_2022}, data operations contain one or more input or output ports.
Each data operation is an atomic operation on the input data to produce certain output data.
In~\cite{moin_model-driven_2022}, each state allows executing functions. 
The keyword \textit{DA\_Preprocess} is used to apply data preparation methods on a specific dataset.
\ins{In~\cite{Ming2023}, feature preparation is given by the possibility to add additional layers and strategies of the image processing layers of the neural network.}

In~\cite{morales_towards_2022}, features can be prepared with specific \textit{feature extraction techniques}, and data can be transformed with data engineering techniques, e.g., \textit{Regression substitution}.
In~\cite{pinedaRADENNDomainSpecificLanguage2023}, feature preparation is given due to the extendability of the build-in functions and due to the fact that datsets can be modified with the build-in functions.
In~\cite{radlerCodeGenerationMachine2023}\hiddencite{removed \cite{radlerModelDrivenEngineeringMethod2023}}, feature preparation is given with the application of stereotypes specifically applied to transform and arrange data attributes.

\subsubsection*{\textbf{Model Training}}
The specification of an algorithm and the related training of the model is depicted in 14 of the 18 approaches.
The types of algorithms can be separated in Inference~\cite{koseler_realization_2019}, Machine Learning~\cite{hartmann_meta-modelling_2019,meacham_adaptivesystems_2021,melchor_model-driven_2022,morales_towards_2022,radlerCodeGenerationMachine2023}\hiddencite{removed \cite{radlerModelDrivenEngineeringMethod2023}} and Deep Learning~\cite{al-azzoni_model_2020,bhattacharjee_stratum_2019,hartsell_model-based_2019,Ming2023,kusmenko_engineering_2019,moin_model-driven_2022,pinedaRADENNDomainSpecificLanguage2023} using Neural Networks.

\paragraph{Inference}
\cite{koseler_realization_2019} extended the approach of~\cite{breuker_towards_2014} with the required implementation using SysML and Papyrus modeling framework.
Within the original approach~\cite{breuker_towards_2014}, model training is given by an assignment for each variable, whether it is an observed variable, a random variable, or a standard variable. 
Details on hyper-parameter tuning are not given.

\paragraph{Machine Learning}
In~\cite{hartmann_next_2017,hartmann_meta-modelling_2019}, various algorithm models can be used with specific input (learning) and output attributes.
In~\cite{meacham_adaptivesystems_2021}, the algorithm (referred to as \textit{approach}) is specified aligned with various hyper-parameters, e.g., Random Forest Cross Validation Folds.
In~\cite{melchor_model-driven_2022}, the algorithm type (e.g., Random Forest) with a specific task type (e.g., Classification) can be described.
Hyper-parameters are not presented in the metamodel.
In~\cite{morales_towards_2022}, hyper-parameters and performance criteria can be specified for each AI model.
In~\cite{radlerCodeGenerationMachine2023}\hiddencite{removed \cite{radlerModelDrivenEngineeringMethod2023}}, algorithm type, parameters and train/test splitting can be applied using stereotypes.

\paragraph{Deep Learning}
In~\cite{al-azzoni_model_2020}, the training is defined using an \textit{MLPDescription} block with certain learning rules like Backpropagation
Further details on other hyper-parameters or the output's facilitation are not given.
In~\cite{bhattacharjee_stratum_2019}, an algorithm for the training is defined in the metamodel.
Moreover, hyper-parameters are defined and applied to a specific algorithm in the editor.
In~\cite{hartsell_model-based_2019}, an experimental model defines the model training.
The details of the implementation can be found in the Jupyter Notebooks.
In~\cite{kusmenko_engineering_2019}, the training of NN is given with possibilities to specify the network layers and connections.
In~\cite{moin_model-driven_2022}, state diagrams are used to define various steps of the algorithm. 
With the state keyword \textit{DA\_Train}, various training-related settings are made, and with \textit{DA\_Predict}, the trained model can be applied to data.
\ins{In~\cite{Ming2023}, the modelling of various layers is given by the extension of the metamodel and the possibility to define and deploy executable code on a simulation model.}

In~\cite{pinedaRADENNDomainSpecificLanguage2023}, the training is defined by creating input, hidden and output layer and executed with various parameter such as Epoch definition.

\subsubsection*{\textbf{Metrics/Evaluation}}
To assess the validity of an algorithm, six of the 18 approaches integrate the modeling of metrics.

In~\cite{hartsell_model-based_2019}, the metrics are applied directly in the Jupyter Notebooks, which is not actually a modeling approach.
Nevertheless, the Jupyter Notebook is integrated into the model.
So it can be considered as part of the model.

In~\cite{bhattacharjee_stratum_2019}, metrics are integrated into the metamodel and can be applied to the training output.
In~\cite{kusmenko_engineering_2019}, the evaluation metrics are selected using the name of the metrics, e.g., Mean Squared Error (MSE).

In~\cite{moin_model-driven_2022}, basic metrics such as Mean Absolute Error (MAE) or MSE can be applied to the algorithms, such as regression algorithms.

In~\cite{pinedaRADENNDomainSpecificLanguage2023}, there are two approaches to apply metrics. For classification cases, average accuracy, macro-average sensitiviy and macro averaging specificity are evaluated. For regression cases the mean-square-error, mean-absolute-error and coefficient of determination are evaluated.


In~\cite{radlerCodeGenerationMachine2023}\hiddencite{removed \cite{radlerModelDrivenEngineeringMethod2023}}, various metrics such as mean-square-error or mean-absolute-error can be applied based on the stereotypes defined.

\subsection{Frameworks (Methods \& Tools)}
In Table~\ref{tab:methodsAndTools}, we present a non-exhaustive list of building blocks (rows) corresponding to the approaches discussed in the selected studies (columns). Our categorization focuses on elements related to MDE and AI/ML. Specifically, we distinguish between language workbenches (LW), related languages (in addition to those proposed by the approaches themselves), and various tools.

Notably, several elements are reused across different approaches. In the MDE domain, language workbenches such as EMF, Xtext, WebGME, and Greycat play a significant role. Meanwhile, in the AI/ML domain, the Python programming language stands out as the most prevalent element.


\definecolor{mygray}{gray}{0.9} 

\newcommand{\toolCat}[1]{%
  \ifx#1\mde
    #1
  \else\ifx#1\ai
    #1
  \else\ifx#1\other
    #1
  \else
    Error: Invalid value for \texttt{\string\myCategory}.
  \fi\fi\fi
}

\newcommand{\mde}{MDE}
\newcommand{\ai}{AI/ML}
\newcommand{\other}{\textit{Other}}

\begin{table}
\centering
\caption{Used Methods and Tools (RQ4)}
\label{tab:methodsAndTools}
\resizebox{\linewidth}{!}{%
\begin{tabular}{|c|c|c|c|c|c|c|c|c|c|c|c|c|c|c|c|c|c|c|c|c|c|c|} 
\hline

\hline
\textbf{Tools or Method} & \textbf{Cat.} & \textbf{Type} & 
\cite{al-azzoni_model_2020} & 
\cite{bergerEVOALDomainSpecificLanguageBased2023} & 
\cite{bhattacharjee_stratum_2019} & 
\cite{de_la_vega_lavoisier_2020} & 
\cite{giner-miguelez_domain-specific_2022} & 
\cite{hartmann_next_2017} & 
\cite{hartmann_meta-modelling_2019} & 
\cite{hartsell_model-based_2019} & 
\cite{Ming2023} & 
\cite{koseler_realization_2019} & 
\cite{kusmenko_engineering_2019} & 
\cite{meacham_adaptivesystems_2021} & 
\cite{melchor_model-driven_2022} & 
\cite{moin_model-driven_2022} & 
\cite{morales_towards_2022}
&\cite{pinedaRADENNDomainSpecificLanguage2023} & 
\cite{radlerCodeGenerationMachine2023} & 
\cite{ries_mde_2021}
& \textit{Total} \\ 
\hline
Alloy &\toolCat{\other}& Lang & & & & & & & & & & & & & & & & & & $\blacksquare$ & 1 \\ 
\hline
BPMN &\toolCat{\mde}& Lang & & & & & & & & & & & & & & & & & & $\blacksquare$ & 1 \\ 
\hline
DeepForge~\cite{hartsell_model-based_2019} &\toolCat{\ai}& Tool & & & & & & & & $\blacksquare$ & & & & & & & & & & & 1 \\ 
\hline
\rowcolor{mygray} EMF &\toolCat{\mde}& LW & $\blacksquare$ & & & $\blacksquare$ & & & & & & & & & $\blacksquare$ & & & & $\blacksquare$ &$\blacksquare$ & 5 \\ 
\hline
Epsilon &\toolCat{\mde}& Tool & $\blacksquare$ & & & & & & & & & $\blacksquare$ & & & & & & & & & 2 \\ 
\hline
GSN~\cite{kelly2004goal} &\toolCat{\mde}& Lang & & & & & & & & $\blacksquare$ & & & & & & & & & & & 1 \\ 
\hline
\rowcolor{mygray} GreyCat &\toolCat{\mde}& LW & & & & & & $\blacksquare$ & $\blacksquare$ & & & & & & & & & & & & 2 \\ 
\hline
\rowcolor{mygray} Jetbrains MPS &\toolCat{\mde}& LW & & & & & & & & & & & & $\blacksquare$ & & & & & & & 1 \\ 
\hline
Jupyter Notebook &\toolCat{\ai}& Tool & & & & & & & & $\blacksquare$ & & & & & & & & & $\blacksquare$ & & 2 \\ 
\hline
\rowcolor{mygray} Langium &\toolCat{\mde}& LW & & & & &$\blacksquare$ & & & & & & & & & & & & & $\blacksquare$& 2 \\ 
\hline
\rowcolor{mygray} MontiCore &\toolCat{\mde}& LW & & & & & & & & & & & $\blacksquare$ & & & & & & & & 1 \\ 
\hline
Papyrus &\toolCat{\mde}& Tool & & & & & & & & & & $\blacksquare$ & & & & $\blacksquare$ & & & $\blacksquare$ & & 3 \\ 
\hline
Python &\toolCat{\ai}& Lang & & & & $\blacksquare$ & & & & & & & & & & & $\blacksquare$ & & $\blacksquare$ 
& & 3 \\ 
\hline
ROSMOD~\cite{rosmod} &\toolCat{\mde}& Tool & & & & & & & & $\blacksquare$ & & & & & & & & & & & 1 \\ 
\hline
SEAM\footnote{\url{https://standards.nasa.gov/standard/nasa/nasa-std-87291}} &\toolCat{\other}& Std & & & & & & & & $\blacksquare$ & & & & & & & & & & & 1 \\ 
\hline
SQL &\toolCat{\other}& Lang & & & & & & & & & & & & & & & & & & $\blacksquare$& 1 \\ 
\hline
Sirius &\toolCat{\mde}& Tool & & & & & & & $\blacksquare$ & & & & & & & & $\blacksquare$ & & & & 2 \\ 
\hline
ThingML~\cite{thingml} &\toolCat{\mde}& Lang & & & & & & & & &$\blacksquare$ & & & & & $\blacksquare$ & & & & & 2 \\ 
\hline
\rowcolor{mygray} WebGME &\toolCat{\mde}& LW & & & $\blacksquare$ & & & & & $\blacksquare$ & & & & & & & & & & & 2 \\ 
\hline
\rowcolor{mygray} Xtext &\toolCat{\mde}& LW & & & & $\blacksquare$ & & & & & & & & & & $\blacksquare$ & & & & $\blacksquare$ & 3 \\
\hline
\textit{Total} && &2 &1 &1 &3 &1 &1 &2 &6 &1 &2 &1 &1 &1 &3 &2 &1 &4 &6& \\
\hline
\end{tabular}
}
\end{table}



\subsection{Availability of Artifacts and Application Domains}

Artifacts serve as a means to facilitate the replication of research results. Table~\ref{tab:artifacts} indicates whether artifacts are referenced in the publication as online resources or are absent altogether. Additionally, the table depicts the type of application mentioned in the publication or inherently provided through the evaluation sample. If no specific domain is explicitly mentioned or derivable, we annotate it as \textit{Unknown}.

\begin{table}
\centering
\caption{Availability and type of artifacts aligned with the type of application}
\label{tab:artifacts}
\resizebox{\linewidth}{!}{%
\begin{tabular}{|l|c|c|c|l|} 
\hline
 {\begin{tabular}[c]{@{}l@{}}\textbf{Selected}\\\textbf{
Study}\end{tabular}} & \multicolumn{1}{l|}{\begin{tabular}[c]{@{}l@{}}\textbf{In the}\\\textbf{Study}\end{tabular}} & \multicolumn{1}{l|}{\begin{tabular}[c]{@{}l@{}}\textbf{Online }\\\textbf{(Git, Server)}\end{tabular}} & \multicolumn{1}{l|}{\textbf{No Artifacts}} & \textbf{Type of Application} \\ 
\hline
\cite{al-azzoni_model_2020} & $\blacksquare$ &  &  & Datasets \\ 
\hline
\cite{bergerEVOALDomainSpecificLanguageBased2023} &  & $\blacksquare$ &  & Datasets \\ 
\hline
\cite{bhattacharjee_stratum_2019} &  & $\blacksquare$ &  & IoT \\ 
\hline
\cite{de_la_vega_lavoisier_2020} &  & $\blacksquare$ &  & Datasets \\ 
\hline
\cite{giner-miguelez_domain-specific_2022} &  & $\blacksquare$ &  & Datasets \\ 
\hline
\cite{hartmann_next_2017} &  & $\blacksquare$ &  & IoT \\ 
\hline
\cite{hartmann_meta-modelling_2019} &  & $\blacksquare$ &  & Unknown \\ 
\hline
\cite{hartsell_model-based_2019} &  & $\blacksquare$ &  & IoT (CPS) \\ 
\hline
\cite{Ming2023} &  & $\blacksquare$ &  & IoT \\ 
\hline
\cite{koseler_realization_2019} &  & $\blacksquare$ & & Datasets \\ 
\hline
\cite{kusmenko_engineering_2019} &  & $\blacksquare$ &  & IoT (Image) \\ 
\hline
\cite{meacham_adaptivesystems_2021} &  &  & $\blacksquare$ & Adaptive Systems \\ 
\hline
\cite{melchor_model-driven_2022} &  & $\blacksquare$ &  & Datasets \\ 
\hline
\cite{moin_model-driven_2022} &  & $\blacksquare$ &  & IoT \\ 
\hline
\cite{pinedaRADENNDomainSpecificLanguage2023} & & $\blacksquare$  &  & Datasets \\ 
\hline
\cite{morales_towards_2022} & $\blacksquare$ &  &  & Unknown \\ 
\hline
\cite{radlerCodeGenerationMachine2023} &  & $\blacksquare$ &  & IoT (Datasets) \\
\hline
\cite{ries_mde_2021} &  & $\blacksquare$ &  & Datasets \\ 
\hline
\end{tabular}
}
\end{table} 


As a result, eight approaches work with datasets that can originate from any domain. The processing of Internet of Things (IoT) data is presented in five approaches, while one approach is more specific to image data.
\section{Discussion}\label{sec:discussion}



\begin{figure*}
    \centering
    \begin{tikzpicture}
    \begin{axis}[
            xlabel=$Year$,     ylabel=$Number\;of\;publications$,
            ymin=0,ymax=7,
            ytick={1,2,3,4,5,6,7},
            yticklabels={1,2,3,4,5,6,7},
            symbolic x coords={2017, 2019, 2020, 2021, 2022, 2023},
            xtick=data]
            
            \addplot[ybar,fill=blue] coordinates {
                (2019,  6)
                (2020,  2)
                (2021,  3)
                (2022,  3)
                (2023,  4)
            };
        \end{axis}  
    \end{tikzpicture}
    \caption{Number of publication over year}
    \label{fig:publications_over_year}
\end{figure*}
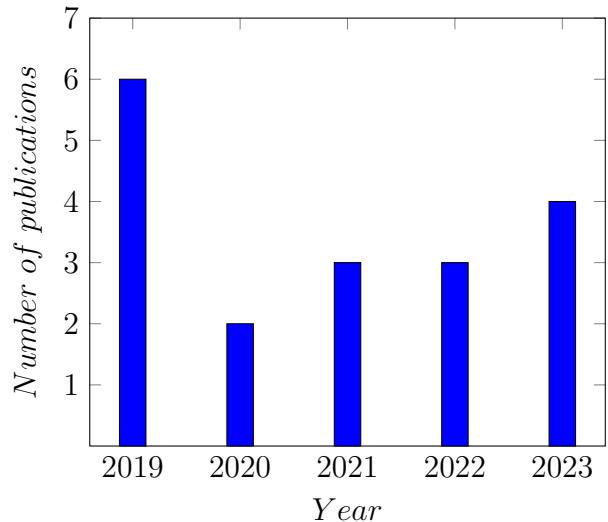

MDE has long been a cornerstone of software engineering research. Meanwhile, AI has witnessed remarkable advancements, with applications spanning various domains. Recently, the convergence of MDE and AI has emerged as a promising research direction.
Figure~\ref{fig:publications_over_year} illustrates the publication trend of selected studies over time, emphasizing the certain interest in MDE4AI.
We observe that the MODELS conference\footnote{\url{https://conf.researchr.org/series/models}} (with three publications) and the MODELSWARD conference\footnote{\url{https://modelsward.scitevents.org/}} (with two publications) stand out as the most representative venues.
However, it is essential to acknowledge that the overall volume of collected publications remains limited. 

This limitation may result from our selective exclusion criteria; our focus on MDE4AI led us to disregard AI4MDE, the complementary perspective.

From a research community perspective, the MDE research community remains keen on exploring this intersection further and establishing a focused community. Notably, the annual MDE Intelligence workshop at MODELS stimulates cross-pollination between MDE and AI.

The rest of the discussion is organized according to the research questions in Section~\ref{sec:RQs}.

\subsection{RQ1 - What language aspects of MDE are addressed in the approaches, e.g., abstract syntax, concrete syntax, metamodel, etc.?}


\begin{itemize}

    \item \textit{Abstract Syntax:} Most MDE approaches explicitly address abstract syntax. It’s a fundamental aspect because it forms the foundation for creating models. The abstract syntax is defined using metamodels and/or grammars, typically through a dedicated tool support of DSL development, a.k.a. language workbenches~\cite{iung_systematic_2020}. 

    \item \textit{Concrete Syntax:} The results highlight that MDE approaches from select studies prefer textual modeling rather than graphical modeling. This preference may be influenced by tooling support, ease of manipulation, and expressiveness. It's essential to highlight a recent trend in MDE: the emergence of blended modeling capabilities. Blended modeling empowers engineers to freely select and switch between various notations for the same DSL. The choice of notation(s) can be influenced by diverse factors, including the complexity of graphical tools and the expertise of modelers.

    \item \textit{Secondary aspects: Validation and Runtime Support:} Model validation ensures that models adhere to defined constraints and rules. Although this aspect is not a prominent focus in the selected studies, it is worth noting that language workbenches (LWs) provide mature support for model validation. Leveraging this support may lead to the development of more mature Domain-Specific Languages (DSLs) and approaches. Runtime support is offered only by Hartmann et al.~\cite{hartmann_next_2017} at a metalanguage level through the GreyCat LW. However, this LW is not as common as other existing LWs. Therefore, we can argue that there is still room for improvement and support in this direction.
 
\end{itemize}

\subsection{RQ2 - Which phases of AI development aligned with the CRISP-DM methodology are covered by the approaches?}

The CRISP-DM development cycle's supported phases are less balanced than the MDE perspectives.
Less than half of the approaches support the early phases, such as business understanding.
Feature preparation is often not mentioned or integrated with only simple features; e.g., normalization of variables is given but not the subsequent processing of pre-processing tasks.
The main focus of the approaches lies in the formalization of model training.
However, most of the approaches only support a small range of algorithms.
Therefore, the applicability might be very case-specific and less flexible.

In summary, it can be seen that multiple approaches depict a specific aspect of the CRISP-DM development cycle, but only a few support more than half of the phases. 

\subsection{RQ3 - Which industrial domains are supported by MDE4AI approaches?}

Most approaches support processing datasets in specific file formats or using data from SQL servers.
Since these datasets can originate from any domain, no focus on a domain can be determined in these approaches.

However, some approaches are rather based on IoT/CPS or sensor data, supporting the integration of production systems or data from the use of e.g., CPS products.
Nevertheless, no domain can be clearly defined here since collecting sensor data is possible in any domain.

\subsection{RQ4 - What are the used methods and MDE tools the proposed approaches rely on?}


The present works are based on a wide variety of tools and methods.
Therefore, the advantages and disadvantages of the individual methods and tools are considered application-dependent, and no statement can be made about the quality of the underlying methods. 
Furthermore, there is a trend towards Eclipse and its products (Papyrus, Sirius, Epsilon, etc).

The adoption of language workbenches (LWs) appears to be crucial. LWs facilitate the creation of textual and graphical concrete syntaxes, essential for model representation and manipulation. Using LWs such as EMF, Xtext, WebGME, and Greycat underscores the importance of standardized tools in Model-Driven Engineering (MDE).

EMF and Xtext can also be identified as state-of-the-art for defining metamodels and grammars or as a basic modeling construct. Adopting other LWs seems more innovative, but it is still limited to a smaller community of researchers.

However, it is worth noting that Greycat seems to be a niche solution, with both uses originating from the same research group. We notice a shift of Greycat features towards being offered as a commercial tool, raising questions about its future adoption within the research community.

MontiCore, a well-known platform in the MDE community~\cite{rumpe_monticore_2017}, is expected to play an increasing role in the intersection between MDE and the AI/ML domain.

Langium, a recent LW based on TypeScript, is particularly suitable for integrating web-based tools and applications, and, therefore, to foster collaborative activities and promote the use of web-based tools with lower installation costs.
Unlike Xtext, Langium does not rely on the Eclipse Modeling Framework (EMF).

\subsection{RQ5 - To what extent is communication between different stakeholders supported by MDE}
Unifying the language of communication can foster communication in an AI project, potentially leading to a better understanding and reduced unknown knowledge among team members.
With less unknown knowledge, unrealistic expectations might be reduced, being one of the categories of why AI projects fail~\cite{westenberger_failure_2022}.
The intersection with other domains is mainly in the initial phases of an AI project, mainly business and data understanding.
Still, the documentation of other phases of the CRISP-DM cycle supports communication among other AI experts.
Only five approaches to interdisciplinary communication support the documentation and integration of business understanding, leading to further research needs.
Data understanding and the downstream processes of the CRISP-DM are more often supported. However, still, further integration of MDE techniques is required due to the early development of some of the approaches.

\subsection{RQ6 - Which challenges and research directions are still open?}\label{sec:futureResearch}

The researchers' observations guide the direction of future research and highlight open challenges. The first observation emphasizes the need for better support in business understanding.

In the literature, experts report that the lack of business values for AI is a challenge. This deficiency may stem from AI experts' limited understanding of specific business contexts. Consequently, AI experts might not recommend realistic and relevant AI approaches.

Aligned with business understanding, project requirements should be formalized. This formalization enables the derivation of project metrics and facilitates assessing the impact of computational support~\cite{rigger_top-down_2019}.

Furthermore, considering that the second-largest group of supported applications in existing works is IoT, integrating systems engineering approaches into MDE4AI could be beneficial. 

Another future work that supports the maturity of MDE4AI is consolidating the advantages of the existing approaches and extending these approaches to fit various use cases.
Combining various approaches to a comprehensive methodology regarding MDE4AI could streamline the research topic and foster the development of MDE4AI toolboxes.

It is worth noting that this study did not identify any valuable contributions, despite the recent trends in low-code and no-code development platforms targeting intelligent systems. One possible reason could be the complexity of the engineering process for AI software systems, which remains too high for platforms aiming at the highest process automation. Consequently, we anticipate more valuable contributions in AI4MDE specifically tailored for low-code/no-code platforms. Such contributions would cater to the needs of citizen developers who can leverage intelligent chatbots (e.g.,~\cite{Xatkit}) to accomplish platform tasks.

Apart from combining the research workforces, future research needs to focus on engineering collaboration using methods designed for collaborative modeling. In this regard, the WebGME and Langium LWs are valuable MDE assets.
Concerning more extensive or interdisciplinary projects, the live or collaborative work on a single model could increase the development performance and the benefit and acceptance of MDE4AI.

Next, the output of MDE4AI is often derived from Python code.
Python is an easy-to-understand, well-known, daily-used language used by AI experts that might lead to changes in the Python code rather than the model.
Consequently, full code generation is not applied, leading to no single source of information because partial truth of information is stored in the model and partial in the Python code~\cite{brambilla_model-driven_2017}.
In this context, it is necessary to elaborate a closed-loop process that feeds the results of the executed algorithm back into the model or adjusts the model in case of changes in the code, e.g., in Python.
With this closed-loop approach, the model is always up-to-date, and further, collaboration with others potentially improves because of the abstract representation of the actual changes.

Finally, only a few approaches mention user studies to assess the impact and benefits of MDE4AI. User studies are required to identify unused potentials and further streamline the development towards a user-centered MDE4AI methodology.

Although the conclusion of this research question suggests many improvements, it is questionable whether all the challenges of MDE4AI can be resolved under a single model, method, or framework. 
Therefore, future work needs to investigate whether a separation of concerns, e.g., based on purpose, AI technology such as machine learning, deep learning, etc., might be more valuable and push the development of MDE in the domain of AI further.
Additionally, this can shape the details on the benefits that MDE can bring to AI.

\section{Threats to Validity}
\label{sec:threatstovalidity}
The study's validity describes the extent to which the results are trustworthy and how biases arising from the subjective views of the researcher are avoided during the analysis. 
Validity must be considered at all stages of a study, and several approaches have been proposed in the literature.
Following~\cite{kitchenham_guidelines_2007}, the following threats to validity are considered:

\begin{itemize}
    \item \textit{Construct Validity: }Construct validity describes the validity of the concept or theory behind the study design such that the results are generalizable~\cite{wohlin_experimentation_2012}. In this SLR, construct validity refers to the potentially subjective analysis of the studies and the different ways in which data extraction is conducted. Following the guidelines~\cite{kitchenham_guidelines_2007}, each study analysis is conducted independently by at least two researchers. If the researchers cannot agree on a conclusion, a third researcher evaluates and discusses the literature until there is no disagreement. In addition, each selected literature was evaluated using the quality criteria suggested by~\cite{long_optimising_2020}. A protocol based on~\cite{kitchenham_guidelines_2007} was defined for performing the extraction protocol, which was discussed by the performing researchers after each step. 
    \item \textit{Internal Validity: } Internal validity describes the causal relationships of the researcher's investigation of whether a factor influences an aspect under study. The particular danger is that a third factor has an unknown effect or side effect. To avoid this danger, the same behavior as for construct validity applies, that more than one researcher assesses the causal relationships. In addition, the tactic suggested by~\cite{kitchenham_guidelines_2007} was followed.
    \item \textit{External Validity: } External validity exists when a finding in the selected literature is of interest to others outside the case under study. In this regard, the SLR uses a quality assessment based on~\cite{long_optimising_2020}, so included papers are published in peer-reviewed. Therefore, third-party investigators pre-assessed the selected studies, and the validity of the initial publication is the responsibility of the external authors.
    \item \textit{Conclusion Validity: } The validity of the conclusion relates to concerns about the reproducibility of the study. The concerns in this paper relate to the possible omission of studies. In this regard, the concerns are mitigated by the carefully applied search strategy using multiple digital libraries in conjunction with the snowballing system as per~\cite{kitchenham_guidelines_2007}. In addition, the researchers followed the detailed search protocol as defined in Section~\ref{sec:method} and applied the quality ratings. However, some concerns might exist due to the interdisciplinary nature of the fields involved and the various definitions of modeling and AI, and the overall complexity of the field. The authors tried to mitigate this problem by discussing the keywords with experts in the field, checking various publication venues in different research areas and clearly defining the naming conventions in the background section.
\end{itemize}

\section{Conclusion}\label{sec:conclusion}
AI is emerging in several disciplines today and has recently attracted the interest of the MDE community, with several workshops being held on the subject.
The development of AI requires several development phases, which potentially can be supported using MDE approaches. 
Currently, the support of AI by MDE is still at an early stage of development.
Therefore, it is necessary to understand the existing approaches to support AI to streamline future research and build on existing knowledge.

We conducted an SLR to investigate the existing body of knowledge in MDE approaches to formalize and define AI applications.
To this end, we followed a rigorous, SLR protocol, selected \total{} approaches, and evaluated them for several dimensions of interest, from MDE and AI.

The result showed that the language engineering perspective of MDE for AI is already mature, and some approaches seem applicable in industrial case studies.
The MDE approaches focus on the training phase of the AI approaches, while time-consuming tasks such as data pre-processing are not considered often.
Additionally, the focus is not on improving communication, collaboration, or understanding of the business processes to be supported, which is reported in the literature as a core problem in AI development projects.
Finally, the review showed that the approaches are case-specific and lack general applicability.

Future work in this research area consists of, among other things, consolidating approaches to combine benefits, expanding approaches to be less case-specific, and adopting a closed-loop process that allows for model-based development that potentially leads to an authoritative source of truth.



 \bibliographystyle{plain} 
 \bibliography{bib/Zotero-WritingRefs}

\begin{thebibliography}{10}

\bibitem{akkiraju_characterizing_2020}
Rama Akkiraju, Vibha Sinha, Anbang Xu, Jalal Mahmud, Pritam Gundecha, Zhe Liu, Xiaotong Liu, and John Schumacher.
\newblock Characterizing {{Machine Learning Processes}}: {{A Maturity Framework}}.
\newblock In Dirk Fahland, Chiara Ghidini, J{\"o}rg Becker, and Marlon Dumas, editors, {\em Business {{Process Management}}}, pages 17--31, {Cham}, 2020. {Springer International Publishing}.

\bibitem{al-azzoni_model_2020}
Issam {Al-Azzoni}.
\newblock Model {{Driven Approach}} for {{Neural Networks}}.
\newblock In {\em 2020 {{International Conference}} on {{Intelligent Data Science Technologies}} and {{Applications}} ({{IDSTA}})}, pages 87--94, 2020.

\bibitem{Atouani2021}
Abdallah Atouani, J\"{o}rg~Christian Kirchhof, Evgeny Kusmenko, and Bernhard Rumpe.
\newblock Artifact and reference models for generative machine learning frameworks and build systems.
\newblock In {\em Proceedings of the 20th ACM SIGPLAN International Conference on Generative Programming: Concepts and Experiences}, GPCE 2021, page 55–68, New York, NY, USA, 2021. Association for Computing Machinery.

\bibitem{azevedo_kdd_2008}
Ana Azevedo and Manuel~Filipe Santos.
\newblock {{KDD}}, {{SEMMA}} and {{CRISP-DM}}: {{A Parallel Overview}}.
\newblock {\em IADIS European Conference Data Mining}, pages 182--185, 2008.

\bibitem{basili_goal_1994}
Victor~R. Basili, Gianluigi Caldiera, and H.~Dieter Rombach.
\newblock The {{Goal Question Metric Approach}}.
\newblock pages 1--10, 1994.

\bibitem{Baumann2022}
Nils Baumann, Evgeny Kusmenko, Jonas Ritz, Bernhard Rumpe, and Moritz~Benedikt Weber.
\newblock Dynamic data management for continuous retraining.
\newblock In {\em Proceedings of the 25th International Conference on Model Driven Engineering Languages and Systems: Companion Proceedings}, MODELS '22, page 359–366, New York, NY, USA, 2022. Association for Computing Machinery.

\bibitem{beihoff_world_2014}
B.~Beihoff, C.~Oster, S.~Friedenthal, Christiaan Paredis, D.~Kemp, H.~Stoewer, D.~Nichols, and J.~Wade.
\newblock A {{World}} in {{Motion}} \textendash{} {{Systems Engineering Vision}} 2025.
\newblock Technical report, {INCOSE}, {San Diego, California}, 2014.

\bibitem{bencomo2019}
Nelly Bencomo, Sebastian G{\"o}tz, and Hui Song.
\newblock Models@ run. time: a guided tour of the state of the art and research challenges.
\newblock {\em Software \& Systems Modeling}, 18:3049--3082, 2019.

\bibitem{bergerEVOALDomainSpecificLanguageBased2023}
Bernhard~J. Berger, Christina Plump, and Rolf Drechsler.
\newblock {{EVOAL}}: {{A Domain-Specific Language-Based Approach}} to {{Optimisation}}.
\newblock In {\em 2023 {{IEEE Congress}} on {{Evolutionary Computation}} ({{CEC}})}, pages 1--10, Chicago, IL, USA, July 2023. IEEE.

\bibitem{bertolini_machine_2021}
Massimo Bertolini, Davide Mezzogori, Mattia Neroni, and Francesco Zammori.
\newblock Machine {{Learning}} for industrial applications: {{A}} comprehensive literature review.
\newblock {\em Expert Systems with Applications}, 175:114820, August 2021.

\bibitem{bhattacharjee_stratum_2019}
Anirban Bhattacharjee, Yogesh Barve, Shweta Khare, Shunxing Bao, Zhuangwei Kang, Aniruddha Gokhale, and Thomas Damiano.
\newblock {{STRATUM}}: {{A BigData-as-a-Service}} for {{Lifecycle Management}} of {{IoT Analytics Applications}}.
\newblock In {\em 2019 {{IEEE International Conference}} on {{Big Data}} ({{Big Data}})}, pages 1607--1612, 2019.

\bibitem{SWEBOK}
Pierre Bourque and Robert Dupuis.
\newblock Software engineering body of knowledge (swebok).
\newblock {\em IEEE Computer Society, EUA}, 2004.

\bibitem{brambilla_model-driven_2017}
Marco Brambilla, Jordi Cabot, and Manuel Wimmer.
\newblock Model-{{Driven Software Engineering}} in {{Practice}}: {{Second Edition}}.
\newblock {\em Synthesis Lectures on Software Engineering}, 3(1):1--207, March 2017.

\bibitem{breuker_towards_2014}
Dominic Breuker.
\newblock Towards {{Model-Driven Engineering}} for {{Big Data Analytics}} - {{An Exploratory Analysis}} of {{Domain-Specific Languages}} for {{Machine Learning}}.
\newblock {\em 47th Hawaii International Conference on System Sciences, HICSS 2014, Waikoloa, HI, USA, January 6-9, 2014}, pages 758--767, 2014.

\bibitem{brunnbauer_idea-ai_2021}
Matthias Brunnbauer, Gunther Piller, and Franz Rothlauf.
\newblock Idea-{{AI}}: {{Developing}} a {{Method}} for the {{Systematic Identification}} of {{AI Use Cases}}.
\newblock August 2021.

\bibitem{brunnbauer_top-down_2022}
Matthias Brunnbauer, Gunther Piller, and Franz Rothlauf.
\newblock Top-{{Down}} or {{Explorative}}? {{A Case Study}} on the {{Identification}} of {{AI Use Cases}}.
\newblock July 2022.

\bibitem{brunton_data-driven_2021}
Steven~L. Brunton, J.~Nathan~Kutz, Krithika Manohar, Aleksandr~Y. Aravkin, Kristi Morgansen, Jennifer Klemisch, Nicholas Goebel, James Buttrick, Jeffrey Poskin, Adriana~W. {Blom-Schieber}, Thomas Hogan, and Darren McDonald.
\newblock Data-{{Driven Aerospace Engineering}}: {{Reframing}} the {{Industry}} with {{Machine Learning}}.
\newblock {\em AIAA Journal}, 59(8):2820--2847, 2021.

\bibitem{bucaioni2022}
Alessio Bucaioni, Antonio Cicchetti, and Federico Ciccozzi.
\newblock Modelling in low-code development: a multi-vocal systematic review.
\newblock {\em Software and Systems Modeling}, 21(5):1959--1981, 2022.

\bibitem{burgueno_guest_2022}
Lola Burgue{\~n}o, Jordi Cabot, Manuel Wimmer, and Steffen Zschaler.
\newblock Guest editorial to the theme section on {{AI-enhanced}} model-driven engineering.
\newblock {\em Software and Systems Modeling}, 21(3):963--965, June 2022.

\bibitem{burgueno_mde_2021}
Loli Burgue{\~n}o, Alexandru Burdusel, S{\'e}bastien G{\'e}rard, and Manuel Wimmer.
\newblock {{MDE Intelligence}} 2019: 1st {{Workshop}} on {{Artificial Intelligence}} and {{Model-Driven Engineering}}.
\newblock In {\em Proceedings of the 22nd {{International Conference}} on {{Model Driven Engineering Languages}} and {{Systems}}}, {{MODELS}} '19, pages 168--169. {IEEE Press}, 2021.

\bibitem{burgueno_contents_2019}
Loli Burgue{\~n}o, Federico Ciccozzi, Michalis Famelis, Gerti Kappel, Leen Lambers, Sebastien Mosser, Richard~F. Paige, Alfonso Pierantonio, Arend Rensink, Rick Salay, Gabriele Taentzer, Antonio Vallecillo, and Manuel Wimmer.
\newblock Contents for a {{Model-Based Software Engineering Body}} of {{Knowledge}}.
\newblock {\em Software and Systems Modeling}, 18(6):3193--3205, December 2019.

\bibitem{burgueno_mde_2021-1}
Loli Burgue{\~n}o, Marouane Kessentini, Manuel Wimmer, and Steffen Zschaler.
\newblock {{MDE Intelligence}} 2021: 3rd {{Workshop}} on {{Artificial Intelligence}} and {{Model-Driven Engineering}}.
\newblock {\em 2021 ACM/IEEE International Conference on Model Driven Engineering Languages and Systems Companion (MODELS-C)}, pages 148--149, 2021.

\bibitem{czarnecki2005overview}
Krzysztof Czarnecki.
\newblock Overview of generative software development.
\newblock In {\em Unconventional Programming Paradigms: International Workshop UPP 2004, Le Mont Saint Michel, France, September 15-17, 2004, Revised Selected and Invited Papers}, pages 326--341. Springer, 2005.

\bibitem{Xatkit}
Gwendal Daniel, Jordi Cabot, Laurent Deruelle, and Mustapha Derras.
\newblock Xatkit: A multimodal low-code chatbot development framework.
\newblock {\em IEEE Access}, 8:15332--15346, 2020.

\bibitem{davey_systems_2022}
Christopher Davey, Sanford Friedenthal, Sky Matthews, David Nichols, Paul Nielsen, Christopher Oster, Taylor Riethle, Garry Roedler, Paul Schreinemakers, Emma Sparks, and Heinz Stoewer.
\newblock Systems {{Engineering Vision}} 2035 \textendash{} {{Engineering Solutions}} for a {{Better World}}.
\newblock Technical report, {INCOSE}, {San Diego, California}, 2022.

\bibitem{de_la_vega_lavoisier_2020}
Alfonso {de la Vega}, Diego {Garc{\'i}a-Saiz}, Marta Zorrilla, and Pablo S{\'a}nchez.
\newblock Lavoisier: {{A DSL}} for increasing the level of abstraction of data selection and formatting in data mining.
\newblock {\em Journal of Computer Languages}, 60:100987, October 2020.

\bibitem{dejanovic2017textx}
Igor Dejanovi{\'c}, Renata Vaderna, Gordana Milosavljevi{\'c}, and {\v{Z}}eljko Vukovi{\'c}.
\newblock Textx: a python tool for domain-specific languages implementation.
\newblock {\em Knowledge-based systems}, 115:1--4, 2017.

\bibitem{delineGlindaSupportingData2021}
Robert~A DeLine.
\newblock Glinda: {{Supporting Data Science}} with {{Live Programming}}, {{GUIs}} and a {{Domain-specific Language}}.
\newblock In {\em Proceedings of the 2021 {{CHI Conference}} on {{Human Factors}} in {{Computing Systems}}}, pages 1--11, Yokohama Japan, May 2021. ACM.

\bibitem{diruscio2022}
Davide Di~Ruscio, Dimitris Kolovos, Juan de~Lara, Alfonso Pierantonio, Massimo Tisi, and Manuel Wimmer.
\newblock Low-code development and model-driven engineering: Two sides of the same coin?
\newblock {\em Software and Systems Modeling}, 21(2):437--446, 2022.

\bibitem{morgan}
Claudio Di~Sipio, Juri Di~Rocco, Davide Di~Ruscio, and Phuong~T Nguyen.
\newblock Morgan: a modeling recommender system based on graph kernel.
\newblock {\em Software and Systems Modeling}, 22(5):1427--1449, 2023.

\bibitem{dogan_machine_2021}
Alican Dogan and Derya Birant.
\newblock Machine learning and data mining in manufacturing.
\newblock {\em Expert Systems with Applications}, 166:114060, March 2021.

\bibitem{emmert-streib_defining_2019}
Frank {Emmert-Streib} and Matthias Dehmer.
\newblock Defining {{Data Science}} by a {{Data-Driven Quantification}} of the {{Community}}.
\newblock {\em Machine Learning and Knowledge Extraction}, 1(1):235--251, March 2019.

\bibitem{espadinha-cruz_review_2021}
P.~{Espadinha-Cruz}, R.~Godina, and E.M.G. Rodrigues.
\newblock A review of data mining applications in semiconductor manufacturing.
\newblock {\em Processes}, 9(2):1--38, 2021.

\bibitem{fahle_systematic_2020}
Simon Fahle, Christopher Prinz, and Bernd Kuhlenk{\"o}tter.
\newblock Systematic review on machine learning ({{ML}}) methods for manufacturing processes \textendash{} {{Identifying}} artificial intelligence ({{AI}}) methods for field application.
\newblock {\em Procedia CIRP}, 93:413--418, 2020.

\bibitem{forootan_machine_2022}
M.M. Forootan, I.~Larki, R.~Zahedi, and A.~Ahmadi.
\newblock Machine {{Learning}} and {{Deep Learning}} in {{Energy Systems}}: {{A Review}}.
\newblock {\em Sustainability (Switzerland)}, 14(8), 2022.

\bibitem{fowler_domain_2010}
Martin Fowler.
\newblock {\em Domain {{Specific Languages}}}.
\newblock {Addison-Wesley Professional}, 1st edition, 2010.

\bibitem{giner-miguelez_domain-specific_2022}
Joan Giner-Miguelez, Abel G\'{o}mez, and Jordi Cabot.
\newblock Describeml: A tool for describing machine learning datasets.
\newblock In {\em Proceedings of the 25th International Conference on Model Driven Engineering Languages and Systems: Companion Proceedings}, MODELS '22, page 22–26, New York, NY, USA, 2022. Association for Computing Machinery.

\bibitem{GIRAY2021}
Görkem Giray.
\newblock A software engineering perspective on engineering machine learning systems: State of the art and challenges.
\newblock {\em Journal of Systems and Software}, 180:111031, 2021.

\bibitem{GoodBengCour16}
Ian~J. Goodfellow, Yoshua Bengio, and Aaron Courville.
\newblock {\em Deep Learning}.
\newblock MIT Press, Cambridge, MA, USA, 2016.
\newblock \url{http://www.deeplearningbook.org}.

\bibitem{thingml}
Nicolas Harrand, Franck Fleurey, Brice Morin, and Knut~Eilif Husa.
\newblock Thingml: a language and code generation framework for heterogeneous targets.
\newblock In {\em Proceedings of the ACM/IEEE 19th international conference on model driven engineering languages and systems}, pages 125--135, 2016.

\bibitem{hartmann_next_2017}
Thomas Hartmann, Assaad Moawad, Francois Fouquet, and Yves Le~Traon.
\newblock The next {{Evolution}} of {{MDE}}: {{A Seamless Integration}} of {{Machine Learning}} into {{Domain Modeling}}.
\newblock {\em Software \& Systems Modeling}, 18:1285--1304, 2019.

\bibitem{hartmann_meta-modelling_2019}
Thomas Hartmann, Assaad Moawad, Cedric Schockaert, Francois Fouquet, and Yves Le~Traon.
\newblock Meta-{{Modelling Meta-Learning}}.
\newblock In {\em 2019 {{ACM}}/{{IEEE}} 22nd {{International Conference}} on {{Model Driven Engineering Languages}} and {{Systems}} ({{MODELS}})}, pages 300--305, 2019.

\bibitem{hartsell_model-based_2019}
Charles Hartsell, Nagabhushan Mahadevan, Shreyas Ramakrishna, Abhishek Dubey, Theodore Bapty, Taylor Johnson, Xenofon Koutsoukos, Janos Sztipanovits, and Gabor Karsai.
\newblock Model-{{Based Design}} for {{CPS}} with {{Learning-Enabled Components}}.
\newblock In {\em Proceedings of the {{Workshop}} on {{Design Automation}} for {{CPS}} and {{IoT}}}, {{DESTION}} '19, pages 1--9, {New York, NY, USA}, 2019. {Association for Computing Machinery}.

\bibitem{henderson_value_2021}
Kaitlin Henderson and Alejandro Salado.
\newblock Value and benefits of model-based systems engineering ({{MBSE}}): {{Evidence}} from the literature.
\newblock {\em Systems Engineering}, 24(1):51--66, January 2021.

\bibitem{Ming2023}
Ming Hu, E.~Cao, Hongbing Huang, Min Zhang, Xiaohong Chen, and Mingsong Chen.
\newblock Aiotml: A unified modeling language for aiot-based cyber–physical systems.
\newblock {\em IEEE Transactions on Computer-Aided Design of Integrated Circuits and Systems}, 42(11):3545--3558, 2023.

\bibitem{huldt_state--practice_2019}
T.~Huldt and I.~Stenius.
\newblock State-of-practice survey of model-based systems engineering.
\newblock {\em Systems Engineering}, 22(2):134--145, March 2019.

\bibitem{iung_systematic_2020}
Aníbal Iung, João Carbonell, Luciano Marchezan, Elder Rodrigues, Maicon Bernardino, Fabio~Paulo Basso, and Bruno Medeiros.
\newblock Systematic mapping study on domain-specific language development tools.
\newblock {\em Empirical Software Engineering}, 25(5):4205--4249, September 2020.

\bibitem{kahani_survey_2019}
Nafiseh Kahani, Mojtaba Bagherzadeh, James~R. Cordy, Juergen Dingel, and Daniel Varró.
\newblock Survey and classification of model transformation tools.
\newblock {\em Software and Systems Modeling}, 18(4):2361--2397, 2019.

\bibitem{kelly2004goal}
Tim Kelly and Rob Weaver.
\newblock The goal structuring notation--a safety argument notation.
\newblock In {\em Proceedings of the dependable systems and networks 2004 workshop on assurance cases}, volume~6. Citeseer Princeton, NJ, 2004.

\bibitem{kitchenham_guidelines_2007}
B.~Kitchenham and S.~Charters.
\newblock Guidelines for performing {{Systematic Literature Reviews}} in {{Software Engineering}}.
\newblock Technical report, 2007.

\bibitem{kitchenham_systematic_2013}
Barbara Kitchenham and Pearl Brereton.
\newblock A systematic review of systematic review process research in software engineering.
\newblock {\em Information and Software Technology}, 55(12):2049--2075, December 2013.

\bibitem{koseler_realization_2019}
Kaan Koseler, Kelsea McGraw, and Matthew Stephan.
\newblock Realization of a {{Machine Learning Domain Specific Modeling Language}}: {{A Baseball Analytics Case Study}}.
\newblock In Slimane Hammoudi, Lu{\'i}s~Ferreira Pires, and Bran Selic, editors, {\em Proceedings of the 7th {{International Conference}} on {{Model-Driven Engineering}} and {{Software Development}}, {{MODELSWARD}} 2019, {{Prague}}, {{Czech Republic}}, {{February}} 20-22, 2019}, pages 13--24. {SciTePress}, 2019.

\bibitem{Kourouklidis2020}
Panagiotis Kourouklidis, Dimitris Kolovos, Nicholas Matragkas, and Joost Noppen.
\newblock Towards a low-code solution for monitoring machine learning model performance.
\newblock In {\em Proceedings of the 23rd ACM/IEEE International Conference on Model Driven Engineering Languages and Systems: Companion Proceedings}, MODELS '20, New York, NY, USA, 2020. Association for Computing Machinery.

\bibitem{rosmod}
Pranav~Srinivas Kumar, William Emfinger, Amogh Kulkarni, Gabor Karsai, Dexter Watkins, Benjamin Gasser, Cameron Ridgewell, and Amrutur Anilkumar.
\newblock Rosmod: a toolsuite for modeling, generating, deploying, and managing distributed real-time component-based software using ros.
\newblock In {\em 2015 International Symposium on Rapid System Prototyping (RSP)}, pages 39--45, 2015.

\bibitem{kusmenko_engineering_2019}
Evgeny Kusmenko, Svetlana Pavlitskaya, Bernhard Rumpe, and Sebastian Stuber.
\newblock On the {{Engineering}} of {{AI-Powered Systems}}.
\newblock In {\em 2019 34th {{IEEE}}/{{ACM International Conference}} on {{Automated Software Engineering Workshop}} ({{ASEW}})}, pages 126--133, {San Diego, CA, USA}, November 2019. {IEEE}.

\bibitem{long_optimising_2020}
Hannah~A Long, David~P French, and Joanna~M Brooks.
\newblock Optimising the value of the critical appraisal skills programme ({{CASP}}) tool for quality appraisal in qualitative evidence synthesis.
\newblock {\em Research Methods in Medicine \& Health Sciences}, 1(1):31--42, September 2020.

\bibitem{lucio_model_2016}
Levi Lúcio, Moussa Amrani, Juergen Dingel, Leen Lambers, Rick Salay, Gehan M.K.~K Selim, Eugene Syriani, and Manuel Wimmer.
\newblock Model transformation intents and their properties.
\newblock {\em Software and Systems Modeling}, 15(3):647--684, 2016.

\bibitem{madni_economic_2019}
Azad Madni and Shatad Purohit.
\newblock Economic {{Analysis}} of {{Model-Based Systems Engineering}}.
\newblock {\em Systems}, 7(1):12, February 2019.

\bibitem{martinez2022}
Silverio Mart{\'\i}nez-Fern{\'a}ndez, Justus Bogner, Xavier Franch, Marc Oriol, Julien Siebert, Adam Trendowicz, Anna~Maria Vollmer, and Stefan Wagner.
\newblock Software engineering for ai-based systems: a survey.
\newblock {\em ACM Transactions on Software Engineering and Methodology (TOSEM)}, 31(2):1--59, 2022.

\bibitem{meacham_adaptivesystems_2021}
Sofia Meacham, Vaclav Pech, and Detlef Nauck.
\newblock {{AdaptiveSystems}}: {{An Integrated Framework}} for {{Adaptive Systems Design}} and {{Development Using MPS JetBrains Domain-Specific Modeling Environment}}.
\newblock {\em IEEE Access}, 9:127973--127984, 2021.

\bibitem{melchor_model-driven_2022}
Fran Melchor, Roberto {Rodriguez-Echeverria}, Jos{\'e}~M. Conejero, {\'A}lvaro~E. Prieto, and Juan~D. Guti{\'e}rrez.
\newblock A {{Model-Driven Approach}} for {{Systematic Reproducibility}} and {{Replicability}} of {{Data Science Projects}}.
\newblock In Xavier Franch, Geert Poels, Frederik Gailly, and Monique Snoeck, editors, {\em Advanced {{Information Systems Engineering}}}, Lecture {{Notes}} in {{Computer Science}}, pages 147--163, {Cham}, 2022. {Springer International Publishing}.

\bibitem{moin_model-driven_2022}
Armin Moin, Moharram Challenger, Atta Badii, and Stephan G{\"u}nnemann.
\newblock A model-driven approach to machine learning and software modeling for the {{IoT}}: {{Generating}} full source code for smart {{Internet}} of {{Things}} ({{IoT}}) services and cyber-physical systems ({{CPS}}).
\newblock {\em Software and Systems Modeling}, 21(3):987--1014, June 2022.

\bibitem{morales_towards_2022}
Sergio Morales, Robert Claris{\'o}, and Jordi Cabot.
\newblock Towards a {{DSL}} for {{AI Engineering Process Modeling}}.
\newblock {\em Product-Focused Software Process Improvement}, 13709:53--60, 2022.

\bibitem{naveed2024}
Hira Naveed, Chetan Arora, Hourieh Khalajzadeh, John Grundy, and Omar Haggag.
\newblock Model driven engineering for machine learning components: A systematic literature review.
\newblock {\em Information and Software Technology}, 169:107423, 2024.

\bibitem{petersen_guidelines_2015}
Kai Petersen, Sairam Vakkalanka, and Ludwik Kuzniarz.
\newblock Guidelines for conducting systematic mapping studies in software engineering: {{An}} update.
\newblock {\em Information and Software Technology}, 64:1--18, August 2015.

\bibitem{pinedaRADENNDomainSpecificLanguage2023}
Israel Pineda, Dustin {Carri{\'o}n-Ojeda}, and Rigoberto {Fonseca-Delgado}.
\newblock {{RADENN}}: {{A Domain-Specific Language}} for the {{Rapid Development}} of {{Neural Networks}}.
\newblock {\em IEEE Access}, 11:86727--86738, 2023.

\bibitem{piorkowski_how_2021}
David Piorkowski, Soya Park, April~Yi Wang, Dakuo Wang, Michael Muller, and Felix Portnoy.
\newblock How {{AI Developers Overcome Communication Challenges}} in a {{Multidisciplinary Team}}: {{A Case Study}}.
\newblock {\em Proceedings of the ACM on Human-Computer Interaction}, 5(CSCW1):1--25, April 2021.

\bibitem{portugal_survey_2016}
Ivens Portugal, Paulo Alencar, and Donald Cowan.
\newblock A {{Survey}} on {{Domain-Specific Languages}} for {{Machine Learning}} in {{Big Data}}.
\newblock 2016.

\bibitem{provost_data_2013}
Foster Provost and Tom Fawcett.
\newblock Data {{Science}} and its {{Relationship}} to {{Big Data}} and {{Data-Driven Decision Making}}.
\newblock {\em Big Data}, 1(1):51--59, March 2013.

\bibitem{radler_survey_2022}
S.~R{\"a}dler and E.~Rigger.
\newblock A {{Survey}} on the {{Challenges Hindering}} the {{Application}} of {{Data Science}}, {{Digital Twins}} and {{Design Automation}} in {{Engineering Practice}}.
\newblock {\em Proceedings of the Design Society}, 2:1699--1708, May 2022.

\bibitem{radlerCodeGenerationMachine2023}
Simon R{\"a}dler, Matthias Rupp, Eugen Rigger, and Stefanie {Rinderle-Ma}.
\newblock Code {{Generation}} for {{Machine Learning}} using {{Model-Driven Engineering}} and {{SysML}}, July 2023.

\bibitem{ries_mde_2021}
Beno{\^i}t Ries, Nicolas Guelfi, and Benjamin Jahic.
\newblock An {{MDE Method}} for {{Improving Deep Learning Dataset Requirements Engineering}} using {{Alloy}} and {{UML}}.
\newblock In Slimane Hammoudi, Lu{\'i}s~Ferreira Pires, Edwin Seidewitz, and Richard Soley, editors, {\em Proceedings of the 9th {{International Conference}} on {{Model-Driven Engineering}} and {{Software Development}}, {{MODELSWARD}} 2021, {{Online Streaming}}, {{February}} 8-10, 2021}, pages 41--52. {SCITEPRESS}, 2021.

\bibitem{rigger_top-down_2019}
Eugen Rigger, Thomas Vosgien, Kristina Shea, and Tino Stankovic.
\newblock A top-down method for the derivation of metrics for the assessment of design automation potential.
\newblock {\em Journal of Engineering Design}, pages 1--31, October 2019.

\bibitem{RODRIGUESDASILVA2015139}
Alberto {Rodrigues da Silva}.
\newblock Model-driven engineering: A survey supported by the unified conceptual model.
\newblock {\em Computer Languages, Systems \& Structures}, 43:139--155, 2015.

\bibitem{rumpe_monticore_2017}
Bernhard Rumpe, Katrin H{\"o}lldobler, and RWTH Aachen, editors.
\newblock {\em {{MontiCore}} 5 Language Workbench}.
\newblock Number Band 32 in Aachener {{Informatik-Berichte}}, {{Software-Engineering}}. {Shaker Verlag}, {Aachen}, edition 2017 edition, 2017.

\bibitem{sahay2020}
Apurvanand Sahay, Arsene Indamutsa, Davide Di~Ruscio, and Alfonso Pierantonio.
\newblock Supporting the understanding and comparison of low-code development platforms.
\newblock In {\em 2020 46th Euromicro Conference on Software Engineering and Advanced Applications (SEAA)}, pages 171--178, 2020.

\bibitem{saltz_crisp-dm_2020}
Jeff Saltz.
\newblock {{CRISP-DM}} is {{Still}} the {{Most Popular Framework}} for {{Executing Data Science Projects}}, November 2020.

\bibitem{sanchez-pinto_big_2018}
L.~Nelson {Sanchez-Pinto}, Yuan Luo, and Matthew~M. Churpek.
\newblock Big {{Data}} and {{Data Science}} in {{Critical Care}}.
\newblock {\em Chest}, 154(5):1239--1248, November 2018.

\bibitem{sarker_machine_2021}
Iqbal~H. Sarker.
\newblock Machine {{Learning}}: {{Algorithms}}, {{Real-World Applications}} and {{Research Directions}}.
\newblock {\em SN Computer Science}, 2(3):160, March 2021.

\bibitem{schoene2019}
Ren\'e Sch\"one, Johannes Mey, Boqi Ren, and Uwe A\ss~mann.
\newblock Bridging the gap between smart home platforms and machine learning using relational reference attribute grammars.
\newblock In {\em 2019 ACM/IEEE 22nd International Conference on Model Driven Engineering Languages and Systems Companion (MODELS-C)}, pages 533--542, 2019.

\bibitem{schroer_systematic_2021}
Christoph Schr{\"o}er, Felix Kruse, and Jorge~Marx G{\'o}mez.
\newblock A {{Systematic Literature Review}} on {{Applying CRISP-DM Process Model}}.
\newblock {\em Procedia Computer Science}, 181:526--534, 2021.

\bibitem{shinde_review_2018}
Pramila~P. Shinde and Seema Shah.
\newblock A {{Review}} of {{Machine Learning}} and {{Deep Learning Applications}}.
\newblock In {\em 2018 {{Fourth International Conference}} on {{Computing Communication Control}} and {{Automation}} ({{ICCUBEA}})}, pages 1--6, {Pune, India}, August 2018. {IEEE}.

\bibitem{someh_overcoming_2020}
Ida Someh, Barbara Wixom, and Angela Zutavern.
\newblock Overcoming {{Organizational Obstacles}} to {{Artificial Intelligence Value Creation}}: {{Propositions}} for {{Research}}.
\newblock pages 5809--5818, January 2020.

\bibitem{trauer_data-driven_2020}
Jakob Trauer, Sebastian {Schweigert-Recksiek}, Luis Onuma, Karsten Spreitzer, Markus M{\"o}rtl, and Markus Zimmermann.
\newblock Data-{{Driven Engineering}} \textendash{} {{Definitions}} and {{Insights}} from an {{Industrial Case Study}} for a {{New Approach}} in {{Technical Product Development}}.
\newblock January 2020.

\bibitem{voelter2013dsl}
Markus Voelter, Sebastian Benz, Christian Dietrich, Birgit Engelmann, Mats Helander, Lennart~CL Kats, Eelco Visser, and GH~Wachsmuth.
\newblock Dsl engineering-designing, implementing and using domain-specific languages.
\newblock 2013.

\bibitem{westenberger_failure_2022}
Jens Westenberger, Kajetan Schuler, and Dennis Schlegel.
\newblock Failure of {{AI}} projects: Understanding the critical factors.
\newblock {\em Procedia Computer Science}, 196:69--76, January 2022.

\bibitem{wirth_crisp-dm_2000}
R{\"u}diger Wirth and Jochen Hipp.
\newblock {{CRISP-DM}}: {{Towards}} a standard process model for data mining.
\newblock {\em Proceedings of the 4th International Conference on the Practical Applications of Knowledge Discovery and Data Mining}, 2000.

\bibitem{wohlin_experimentation_2012}
Claes Wohlin, Per Runeson, Martin H{\"o}st, Magnus~C. Ohlsson, Bj{\"o}rn Regnell, and Anders Wessl{\'e}n.
\newblock {\em Experimentation in {{Software Engineering}}}.
\newblock {Springer Science \& Business Media}, June 2012.

\bibitem{xu_review_2022}
J.~Xu, M.~Kovatsch, D.~Mattern, F.~Mazza, M.~Harasic, A.~Paschke, and S.~Lucia.
\newblock A {{Review}} on {{AI}} for {{Smart Manufacturing}}: {{Deep Learning Challenges}} and {{Solutions}}.
\newblock {\em Applied Sciences (Switzerland)}, 12(16), 2022.

\end{thebibliography}





\end{document}